# Anisotropic Third Harmonic Generation in Two-Dimensional Tin Sulfide

*George Miltos Maragkakis[\*], Sotiris Psilodimitrakopoulos[\*], Leonidas Mouchliadis, Abdus Salam Sarkar, Andreas Lemonis, George Kioseoglou, and Emmanuel Stratakis[\*]*

G. M. Maragkakis, S. Psilodimitrakopoulos, L. Mouchliadis, A. S. Sarkar, A. Lemonis, G. Kioseoglou, E. Stratakis

Institute of Electronic Structure and Laser Foundation for Research and Technology-Hellas, Heraklion, Crete, 71110, Greece

E-mails: gmaragkakis@physics.uoc.gr; sopsilo@iesl.forth.gr; stratak@iesl.forth.gr

G. M. Maragkakis, E. Stratakis

Department of Physics, University of Crete, Heraklion, Crete, 71003, Greece

G. Kioseoglou

Department of Materials Science and Technology, University of Crete, Heraklion, Crete, 71003, Greece

**Abstract**: The in-plane anisotropic properties of two-dimensional (2D) group IV monochalcogenides provide an additional degree of freedom which can be used in future optoelectronic devices. Here, it is shown that the third harmonic generation (THG) signal produced by ultrathin tin (II) sulfide (SnS) is in-plane anisotropic with respect to the incident linear polarization of the laser field. We fit the experimental polarization-resolved THG (P-THG) measurements with a nonlinear optics model, which accounts for the orthorhombic crystal structure of 2D SnS. We calculate the relative magnitudes of the $\chi^{(3)}$ tensor components by recording and simultaneously fitting both orthogonal components of the P-THG intensity. Furthermore, we introduce a THG anisotropy ratio, whose calculated values compare the total THG intensity when the excitation linear polarization is along the armchair crystallographic direction with the case when it is along the zigzag direction. Our results provide quantitative information on the anisotropic nature of the THG process in SnS, paving the way to a better understanding of anisotropic nonlinear light-matter interactions, and the development of polarization-sensitive nonlinear optical devices.

# 1. Introduction

The nonlinear optical properties of 2D layered materials have been recently attracting considerable interest for both fundamental studies and technological applications, based on light generation of additional frequencies and their modulation, that can also be used for material characterization.[1-6] Second harmonic generation (SHG) and THG by 2D group-VI transition metal dichalcogenides (TMD) crystals, such as $MoS_2$, $WS_2$, $MoSe_2$ and $WSe_2$, have already been at the center of this interest, providing useful information on their structural properties.[1-12] For example, THG microscopy has enabled rapid visualization of grain boundaries in monolayer $MoS_2$.[11]

Recently, another family of layered 2D materials has been gaining growing attention, namely the group-IV monochalcogenides, also known as group IV-VI metal monochalcogenides, and denoted by MX with M = Sn or Ge and X = S or Se.[13-15] MXs are layered, orthorhombic, semiconducting 2D materials, known as phosphorene analogues,[13-16] given that they share similar puckered or wavy lattice structures with phosphorene, a 2D format of black phosphorus.[17,18] Importantly, MXs feature in-plane anisotropic physical properties,[13-15,19,20] originating from their in-plane structural anisotropy, with puckered structure along the armchair (AC) crystallographic direction.[15] Properties with reported in-plane anisotropic response include carrier mobility,[19] optical absorption, reflection, extinction, and refraction,[20] and Raman spectral behavior.[13] The in-plane anisotropic response is exhibited along the distinguished in-plane AC and zigzag (ZZ) crystallographic directions, offering an additional degree of freedom in manipulating their properties.[13-15,19,20] For example, polarization-sensitive photodetectors have been presented, based on the intrinsic linear dichroism of GeSe,[20] and black phosphorus.[21] Furthermore, monolayer MXs are predicted to be multiferroic with coupled ferroelectricity and ferroelasticity, and large spontaneous polarization.[22,23] Indeed, in-plane ferroelectricity has been demonstrated for monolayer SnS at room temperature.[24] Moreover, access to the valley-related degree of freedom has been reported.[25,26]

Recently, studies on the nonlinear optical properties of MXs have also been generating interesting research.[24,27-33] In particular, monolayer MXs have been theoretically predicted to produce giant optical SHG.[27,28] Moreover, polarized SHG spectroscopy on monolayer SnS,[24] efficient and anisotropic SHG in few-layer SnS,[29] SHG imaging of ultrathin SnS,[30] and wavelength-dependent SHG from few-layer ferroelectric SnS,[31] have been reported. Finally, the nonlinear optical absorption properties of SnS,[32] and SnSe,[33] have been studied, revealing their saturable absorption properties.

However, the THG properties of 2D SnS remain unexplored. THG is a process in which three incident photons with frequency ω generate coherent radiation with frequency 3ω.[34] Notably, THG, in contrast to SHG, does not require non-centrosymmetry, and thus can also be observed in centrosymmetric crystals. Importantly, polarization-dependent anisotropic THG has been reported for various 2D materials, including germanium selenide (GeSe),[35] which belongs to the family of MXs, black phosphorus (BP),[36-38] silicon phosphide (SiP),[39] germanium arsenide (GeAs),[40] arsenic trisulfide (As$_2$S$_3$),[41] and rhenium disulfide (ReS$_2$).[42]

In this work, we investigate the P-THG process in ultrathin SnS, produced via liquid phase exfoliation (LPE),[30,43-47] and characterized with various techniques to contain monolayer and bilayer crystals (see Experimental Section). Our methodology is based on nonlinear optical imaging, which has been recently demonstrated as a powerful tool to explore the properties of 2D materials.[30,48-53] With respect to the rotating orientation of the excitation linear polarization, the THG signal is found to be in-plane anisotropic. By using a polarizing beam splitter in front of two orthogonally placed detectors, we simultaneously record the intensity of the parallel and perpendicular polarization components of the THG signal. We then simultaneously fit these two sets of experimental P-THG measurements to the theoretical model, obtaining a single set of parameter values, allowing us to calculate the relative magnitudes of the $\chi^{(3)}$ tensor components. As we demonstrate, this approach provides increased precision and decreased ambiguity in the above calculation,[54] given the fact that the theoretical model contains five free parameters, and thus, several combinations of values could fit into the model. Indeed, the extraction of the third-order coefficients has been reported in literature as a challenging endeavour, in both 2D materials,[37] and bio-tissues.[55] We also introduce and calculate a THG anisotropy ratio, which compares the total THG intensity when the excitation linear polarization is along the AC direction, to the THG intensity when the polarization is along the ZZ direction. All the above analysis is performed for different 2D SnS crystals belonging in the same field of view. By using laser raster-scanning and the acquisition of spatially resolved THG intensities, forming images, we obtain the means of direct comparison regarding the anisotropic nonlinear optical response between different SnS crystals. Besides this, in case of 2D MXs crystals of larger size, this technique can additionally offer large-area information on the presence of grain boundaries and defects.[11,48,49] The demonstrated technique is all-optical, minimally invasive and rapid, and can become a useful tool towards fundamental studies and optoelectronic applications of 2D materials with in-plane anisotropy.

## 2. Results and Discussion

### 2.1. Theoretical Formulation of P-THG in 2D SnS

The crystal structure of 2D SnS is schematically illustrated in **Figure 1**a.[56] In order to describe the interaction of the laser excitation field with an orthorhombic MX crystal, and the generation of the third harmonic field, we employ the Jones formalism.[30,48-53,57] We particularly consider two coordinate systems: the laboratory frame $(X, Y, Z)$, and that defined by the crystal plane $(x, y, z)$, where $z \parallel Z$ (Figure 1b). The laser beam propagates along $Z$-axis, normally incident to the crystal, and is linearly polarized along the sample plane, oriented at an angle $\varphi$ with respect to $X$-axis. In the experiment, the angle $\varphi$ is controlled via a rotating half-waveplate. The $x$-axis is considered to be parallel to the AC crystallographic direction and oriented at angle $\theta$ with respect to $X$-axis (Figure 1b).

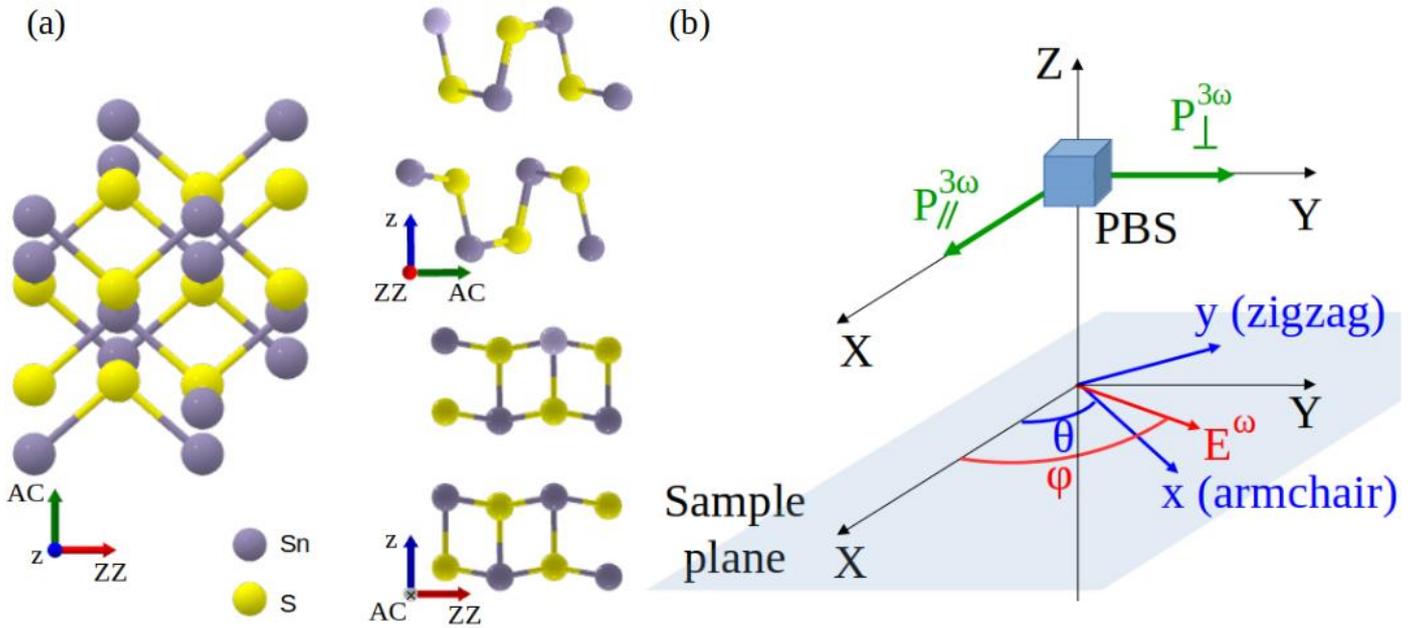

**Figure 1.** a) Schematic illustration of the crystal structure of orthorhombic 2D SnS.[56] b) Schematic illustration of the two coordinate systems adopted in our experimental configuration: the laboratory one ($X$, $Y$, $Z$) and the crystal one ($x$, $y$, $z$), where $z$ and $Z$ coincide. The angles $\varphi$ and $\theta$ describe the orientation of the laser field, and the AC crystallographic direction, relative to $X$ laboratory axis, respectively. In the detection path, a polarizing beam splitter (PBS) analyzes the generated THG field into its two orthogonal components, $P_\parallel^{3\omega}$ and $P_\perp^{3\omega}$.

In our system, the excitation field past the half-wave retardation plate can be expressed in laboratory coordinates by the Jones vector $\begin{pmatrix} E_0 \cos \varphi \\ E_0 \sin \varphi \end{pmatrix}$, where $E_0$ is the amplitude of the electric field. The expression of this vector in crystal coordinates can be derived by multiplying the excitation field with the rotation matrix $\begin{pmatrix} \cos \theta & \sin \theta \\ -\sin \theta & \cos \theta \end{pmatrix}$, giving $E^\omega = \begin{pmatrix} E_0 \cos(\varphi - \theta) \\ E_0 \sin(\varphi - \theta) \end{pmatrix}$.

The 2D MXs belong to the orthorhombic point group C$_{2v}$ (mm2),[13,24,27] and their contracted third-order nonlinear susceptibility tensor, $\chi^{(3)}$, features the following non-zero elements: $\chi_{11}, \chi_{16}, \chi_{18}, \chi_{22}, \chi_{24}, \chi_{29}, \chi_{33}, \chi_{35}, \chi_{37}$.[34,35,58] In this notation, the first subscript refers to: 1: $x$, 2: $y$, 3: $z$, and the second to: 1: $xxx$, 2: $yyy$, 3: $zzz$, 4: $yzz$, 5: $yyz$, 6: $xzz$, 7: $xxz$, 8: $xyy$, 9: $xxy$, 0: $xyz$. As a result, the THG equation is expressed as:[34,35,58]

$$\begin{pmatrix} P_x^{3\omega} \\ P_y^{3\omega} \\ P_z^{3\omega} \end{pmatrix} = \varepsilon_0 \begin{pmatrix} \chi_{11} & 0 & 0 & 0 & 0 & \chi_{16} & 0 & \chi_{18} & 0 & 0 \\ 0 & \chi_{22} & 0 & \chi_{24} & 0 & 0 & 0 & 0 & \chi_{29} & 0 \\ 0 & 0 & \chi_{33} & 0 & \chi_{35} & 0 & \chi_{37} & 0 & 0 & 0 \end{pmatrix} \begin{pmatrix} E_x^\omega E_x^\omega E_x^\omega \\ E_y^\omega E_y^\omega E_y^\omega \\ E_z^\omega E_z^\omega E_z^\omega \\ 3E_y^\omega E_z^\omega E_z^\omega \\ 3E_y^\omega E_y^\omega E_z^\omega \\ 3E_z^\omega E_z^\omega E_x^\omega \\ 3E_x^\omega E_x^\omega E_z^\omega \\ 3E_y^\omega E_y^\omega E_x^\omega \\ 3E_x^\omega E_x^\omega E_y^\omega \\ 6E_x^\omega E_y^\omega E_z^\omega \end{pmatrix} \quad (1)$$

Considering that the pump laser beam is polarized along the sample plane, i.e., $E_z^\omega = 0$, Equation 1 is reduced to:

$$\begin{pmatrix} P_x^{3\omega} \\ P_y^{3\omega} \end{pmatrix} = \varepsilon_0 \begin{pmatrix} \chi_{11}(E_x^\omega)^3 + 3\chi_{18}E_x^\omega(E_y^\omega)^2 \\ \chi_{22}(E_y^\omega)^3 + 3\chi_{29}E_y^\omega(E_x^\omega)^2 \end{pmatrix} \quad (2)$$

and therefore, only terms including four independent $\chi^{(3)}$ elements survive. $\chi_{11}$ and $\chi_{22}$ are known as the on-axis $\chi^{(3)}$ elements,[41] where the row number equals the column number, and they represent contributions to the THG signal when all three incident fields are parallel to $x$-axis (AC direction) or $y$-axis (ZZ direction), respectively. On the other hand, $\chi_{18}$ and $\chi_{29}$ are known as the off-axis elements,[41] where the row number is not equal to the column number, and they represent contributions to the THG signal when the incident fields are not all parallel to each other.

We then transform Equation 2 back to laboratory coordinates. By using a polarizing beam splitter in front of the detectors, we collect the two orthogonal components of the THG field, $P_\parallel^{3\omega} = P_X^{3\omega}$ and $P_\perp^{3\omega} = P_Y^{3\omega}$. The intensities of these two orthogonal THG components, $I_\parallel^{3\omega} = |P_\parallel^{3\omega}|^2$ and $I_\perp^{3\omega} = |P_\perp^{3\omega}|^2$, are calculated to be:

$$I_\parallel^{3\omega} = a\left[(\cos^2(\varphi - \theta) + 3b\sin^2(\varphi - \theta))\cos\theta\cos(\varphi - \theta) - (c\sin^2(\varphi - \theta) + 3d\cos^2(\varphi - \theta))\sin\theta\sin(\varphi - \theta)\right]^2 \quad (3)$$

$$I_\perp^{3\omega} = a\left[(\cos^2(\varphi - \theta) + 3b\sin^2(\varphi - \theta))\sin\theta\cos(\varphi - \theta) + (c\sin^2(\varphi - \theta) + 3d\cos^2(\varphi - \theta))\cos\theta\sin(\varphi - \theta)\right]^2 \quad (4)$$

where:

$$b = \chi_{18}/\chi_{11}, \; c = \chi_{22}/\chi_{11}, d = \chi_{29}/\chi_{11} \quad (5)$$

are the relative magnitudes of the $\chi^{(3)}$ tensor components, and $\alpha$ is a multiplication factor that depends on the square of the $\chi^{(3)}$ tensor element $\chi_{11}$, and the amplitude of the electric field. Details on the above calculation are presented in section S1 in Supporting Information.

The total P-THG intensity can then be obtained through:

$$I^{3\omega} = I_{\parallel}^{3\omega} + I_{\perp}^{3\omega} \quad (6)$$

or equivalently:

$$I^{3\omega} = a\{cos^2(\varphi-\theta)[cos^2(\varphi-\theta) + 3bsin^2(\varphi-\theta)]^2 + sin^2(\varphi-\theta)[csin^2(\varphi-\theta) + 3dcos^2(\varphi-\theta)]^2\} \quad (7)$$

We then present numerical simulations of the theoretical P-THG intensities $I_{\parallel}^{3\omega}$ and $I_{\perp}^{3\omega}$, produced by a 2D MX, described by Equation 3 and 4, respectively. In particular, in Figure S1-S4, Supporting Information, we plot these P-THG modulations in polar diagrams, as function of the orientation of the linearly polarized excitation angle $\varphi$, for different values of the parameters of the model, namely the relative magnitudes of the $\chi^{(3)}$ tensor components $b$, $c$ and $d$ (Figure S1-S3, Supporting Information), and the AC direction $\theta$ (Figure S4, Supporting Information). The two P-THG intensity components are predicted to exhibit two-lobe patterns. It is noted that for specific values of these parameters, a change from a two-lobe to a four-lobe pattern could, in principle, be observed (Figure S6, Supporting Information). Furthermore, we find that for different values of these parameters, the shapes of the P-THG modulations change. Considering that $I_{\parallel}^{3\omega}$ and $I_{\perp}^{3\omega}$ are found to be sensitive to changes of the values of the relative magnitudes of the $\chi^{(3)}$ tensor components $b$, $c$ and $d$, we conclude that we can calculate the values of these quantities, upon measuring the two P-THG intensity components and subsequently fitting to the above nonlinear optics model.

Additionally, we have estimated the smallest detectable change in the AC crystallographic direction of a 2D SnS crystal, based on the P-THG measurements. For this purpose, we have performed numerical simulations using "ideal" P-THG data, subject only to Poisson noise, which is commonly used to describe the effect of photon noise in our detectors, namely photomultiplier tubes.[48] See Figure S5 and S6, and the relevant discussion in section S3 in Supporting Information

The dependency of the two P-THG intensity components to the parameters of the model, namely $\theta$, $b$, $c$ and $d$, constitutes a direct link between the P-THG properties and the in-plane anisotropy of orthorhombic MXs. To account for such anisotropy, we introduce a THG anisotropy ratio (AR), which compares the total THG intensities for incident field polarization along the AC and ZZ crystallographic directions. For $\varphi = \theta$, i.e., when the linear polarization of the incident field coincides with the AC direction, we obtain $I_{AC}^{3\omega} \sim \chi_{11}^2$. On the other hand, for $\varphi = \theta + 90^o$, i.e., when the linear polarization of the incident field coincides with the ZZ direction, we obtain $I_{ZZ}^{3\omega} \sim \chi_{22}^2$. Therefore, we have:

$$AR = \frac{I_{ZZ}^{3\omega}}{I_{AC}^{3\omega}} = \left(\frac{\chi_{22}}{\chi_{11}}\right)^2 = c^2 \quad (8)$$

an expression that correlates the AR with the $\chi^{(3)}$ tensor ratio $c$, which is defined in Equation 5.

We also note that the produced THG field is also linearly polarized at an angle $\theta_{THG}$ given by:

$$\theta_{THG} = \tan^{-1}\left\{\frac{\chi_{22}\tan^2(\varphi - \theta) + 3\chi_{29}}{\chi_{11} + 3\chi_{18}\tan^2(\varphi - \theta)} \tan(\varphi - \theta)\right\} \quad (9)$$

Consequently, by using Equation 9, we could, in principle, calculate the direction $\theta_{THG}$ of the linear polarization of the THG field.

## 2.2. Experimental P-THG Imaging

In **Figure 2**, we present a schematic illustration of the experimental setup, used for P-THG imaging of ultrathin SnS (see Experimental Section for a detailed description of the setup, as well as the sample preparation and characterization). In particular, a fs laser beam is focused onto the sample under study, while a pair of galvanometric mirrors is used for laser raster-scanning over the sample area. We record spatially resolved THG images while rotating the angle of the linear polarization of the fundamental beam. In the detection path, we have used a polarizing beam splitter cube, coupled with two orthogonally placed detectors. Based on this experimental configuration, we simultaneously record the intensity of the two orthogonal THG field components, in a single-shot experiment.

In the inset in Figure 2, we present the power-law dependence of the THG intensity produced by an ultrathin SnS crystal, as function of the excitation laser power. The slope 3, in the double log-scale plot, confirms the THG process.[35-42]

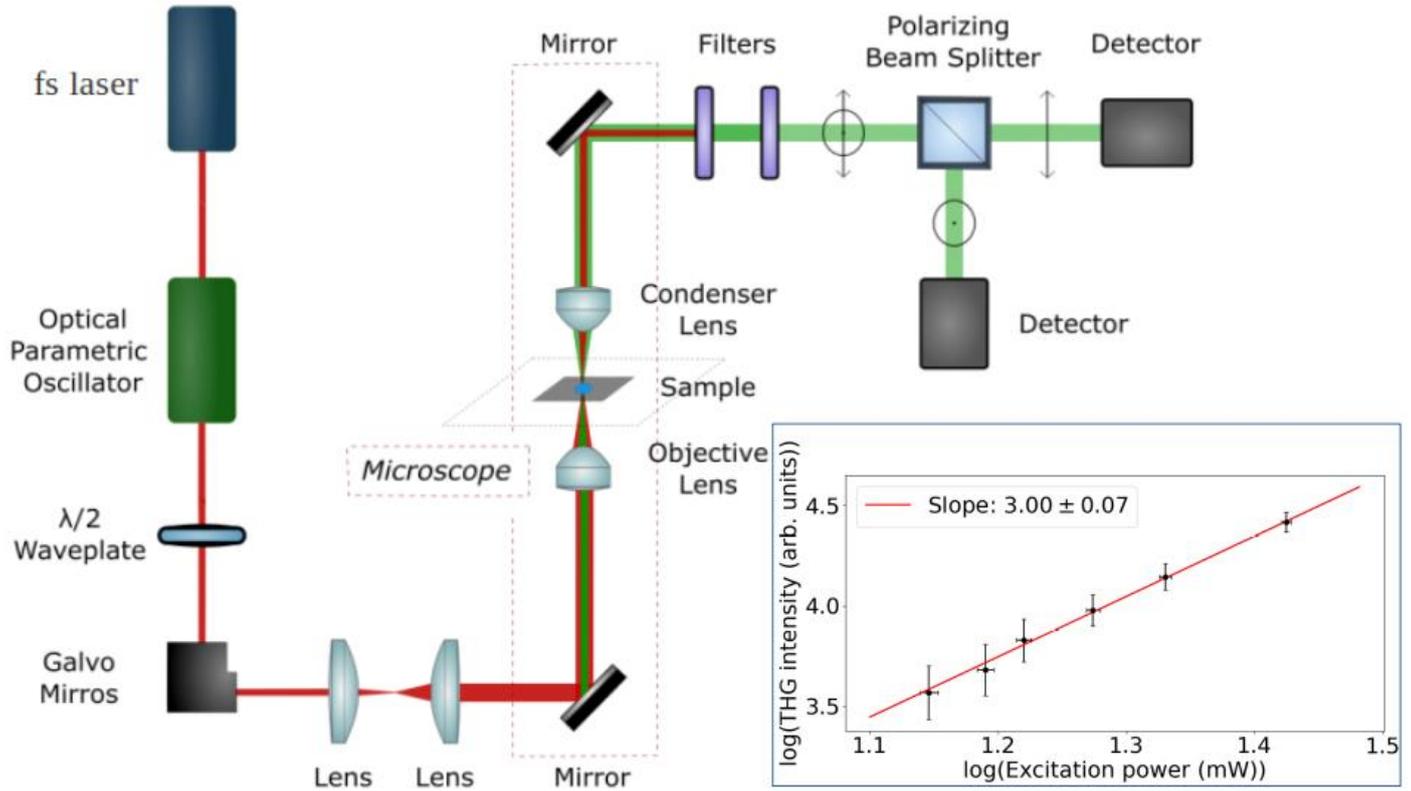

**Figure 2.** Schematic illustration of the nonlinear optical setup, which is based on a fs laser beam coupled to a microscope. The fundamental, 1028 nm, pulses are converted to 1542 nm, by means of an optical parametric oscillator. 2D SnS crystals, excited by the laser beam, generate third harmonic radiation centered at 514 nm. The THG signal is recorded while changing the angle of the linear polarization of the excitation beam, via a rotating half-wave plate, performing P-THG imaging. A pair of galvanometric mirrors enables raster-scanning of stationary SnS crystals, obtaining THG images of the sample area. The THG signal passes through a polarizing beam splitter, allowing the simultaneous measurement of the intensity of the two orthogonal components of the THG field. Inset: Log-scale plot of the THG intensity, produced by an ultrathin SnS crystal, as function of the incident pump power. Black points with the error bars represent the experimental data, and the red line the linear fitting.

In **Figure 3**, we present a series of images of the intensity of the parallel THG component, $I_\parallel^{3\omega}$, produced by ultrathin SnS crystals in the same field of view, recorded while rotating the pump linear polarization angle $\varphi$, indicated by the orange arrow. As can be observed, the THG signals modulate upon $\varphi$ variation. Video S1 ($I_\parallel^{3\omega}$) and S2 ($I_\perp^{3\omega}$), Supporting Information, show all ninety THG images recorded, with $\varphi \in [0°, 360°)$ with step of 4°. We mark four regions of interest (ROIs) containing ultrathin SnS crystals. Additionally, we have studied two more similar crystals in a different field of view, namely ROIs 5 and 6 shown in Figure S7, Supporting Information, for a total sample of six ultrathin SnS crystals under study.

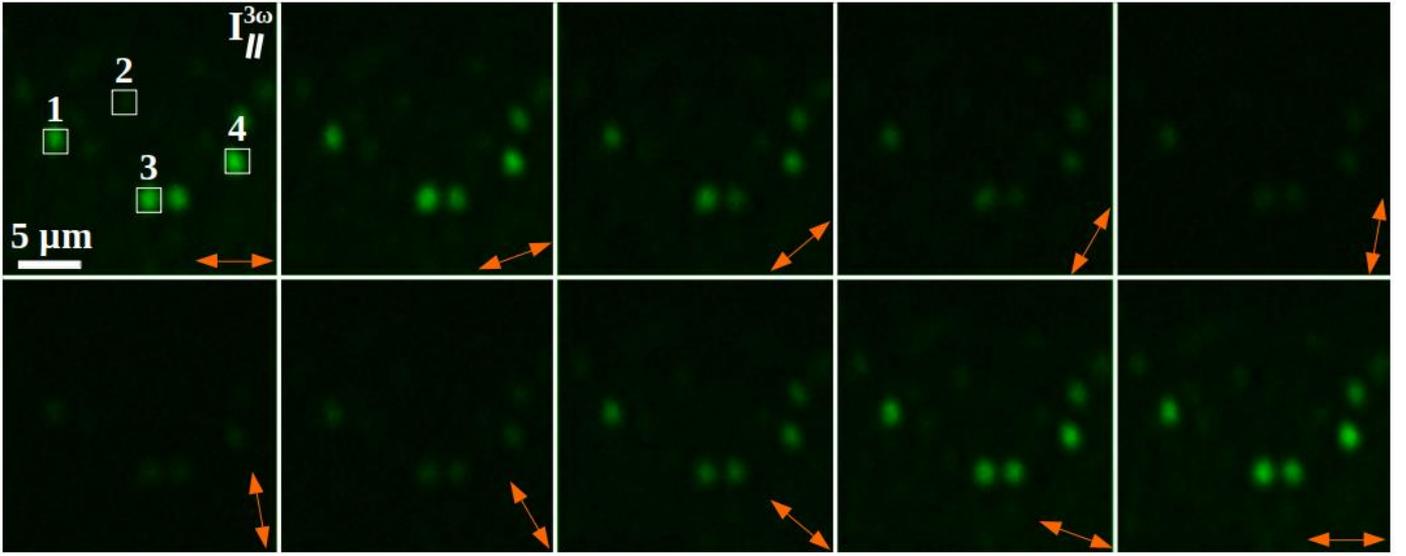

**Figure 3.** Experimental P-THG images of ultrathin SnS crystals belonging in the same field of view. Particularly, we present the intensity of the parallel component of the THG field, $I_{\parallel}^{3\omega}$, for different values of the orientation $\varphi$ of the linearly polarized excitation field, illustrated by the orange arrows. The value of $\varphi$ is varied from 0° to 180° with step of 20°. We mark four ROIs corresponding to ultrathin SnS crystals. Brighter color indicates higher THG intensity.

In **Figure 4**a,b, we present, for the field of view shown in Figure 3, the sum of all ninety collected THG intensity images of $I_{\parallel}^{3\omega}$ and $I_{\perp}^{3\omega}$, respectively. In Figure 4c,d, we plot, in polar diagrams, the modulation of $I_{\parallel}^{3\omega}$ and $I_{\perp}^{3\omega}$, respectively, for the SnS crystals in ROIs 1 and 3, as function of the angle $\varphi$ of the linear polarization of the excitation beam. The two P-THG intensity components are found to exhibit two-lobe patterns, complying with the theoretical prediction. Furthermore, it is evident that these two SnS crystals produce different P-THG modulations (the one is rotated with respect to the other), in accordance with the theoretical prediction. This difference is the signature of nonlinear optical anisotropy between the two crystals.

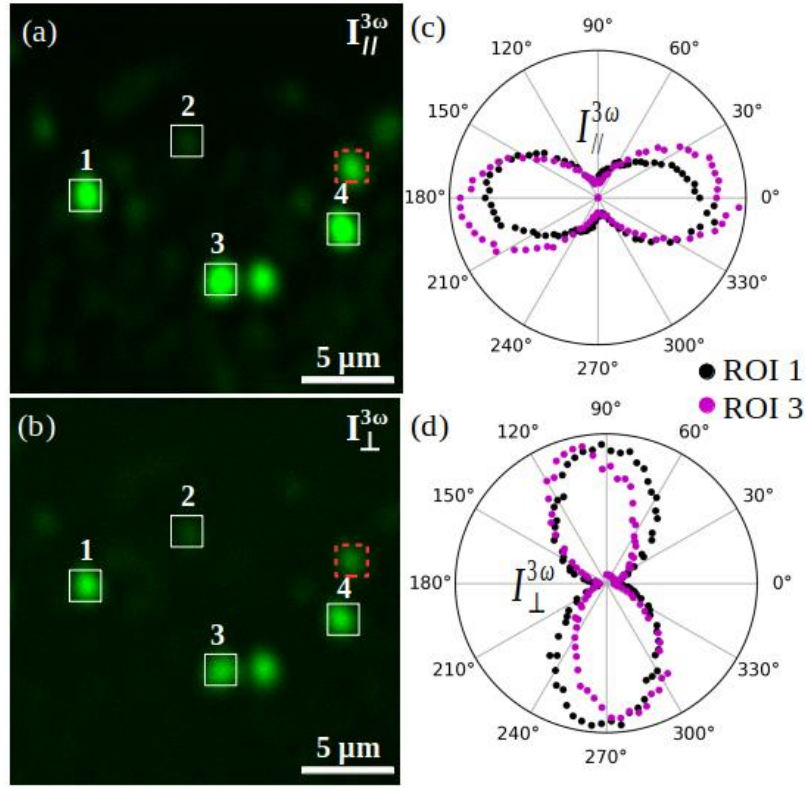

**Figure 4.** Sum of a) $I_{\parallel}^{3\omega}$ and b) $I_{\perp}^{3\omega}$ P-THG intensities, for all orientations $\varphi$ of the excitation linear polarization, corresponding to the field of view shown in Figure 3. We mark the four ROIs we have focused in our analysis, corresponding to ultrathin SnS crystals present in the same field of view. Brighter color indicates higher THG intensity. Experimental polar plots of c) $I_{\parallel}^{3\omega}$ and d) $I_{\perp}^{3\omega}$ P-THG intensities as function of the angle $\varphi$, comparing the P-THG modulations in ROI 1 (in black) and ROI 3 (in magenta).

## 2.3. Experimental Fitting Analysis

We then fit the P-THG experimental data with our theoretical model, to extract the relative magnitudes of the $\chi^{(3)}$ tensor components, *b*, *c*, and *d*. It is important to note that fitting only one P-THG intensity component (either $I_{\parallel}^{3\omega}$ or $I_{\perp}^{3\omega}$) is not sufficient to obtain reliable fitting results. This is clearly demonstrated in Figure S8, Supporting Information, where, for ROI 3, we have fitted only one intensity component, namely $I_{\perp}^{3\omega}$ with Equation 4, and then tested the fitting values to the other component, namely $I_{\parallel}^{3\omega}$ through Equation 3. It is observed that although there is a set of parameter values that fit very well $I_{\perp}^{3\omega}$ (quality of fitting 99.3%), the same set does not fit $I_{\parallel}^{3\omega}$ (quality of fitting 32.2%), producing a four-lobe pattern instead of a two-lobe one. In this work, we have resolved this issue, by simultaneously fitting both $I_{\parallel}^{3\omega}$ and $I_{\perp}^{3\omega}$, obtaining one set of parameter values, with quality of fitting, $R^2$, equal to the average ($R^2 = (R_{\parallel}^2 + R_{\perp}^2)/2$). It is noted that all of the experimental fitting results presented here exhibit average quality of fitting larger than 85%. We also note that the whole procedure of loading and simultaneously fitting both experimental datasets, for an individual crystal, requires less than one minute, rendering our methodology a rapid characterization tool (see also

Experimental Section for the data analysis). Following the above approach, in **Figure 5** and S9, Supporting Information, we present the simultaneous fitting to both P-THG intensity components, for all six ultrathin SnS crystals under study. The agreement between the experimental data and the theoretical fitting is remarkable.

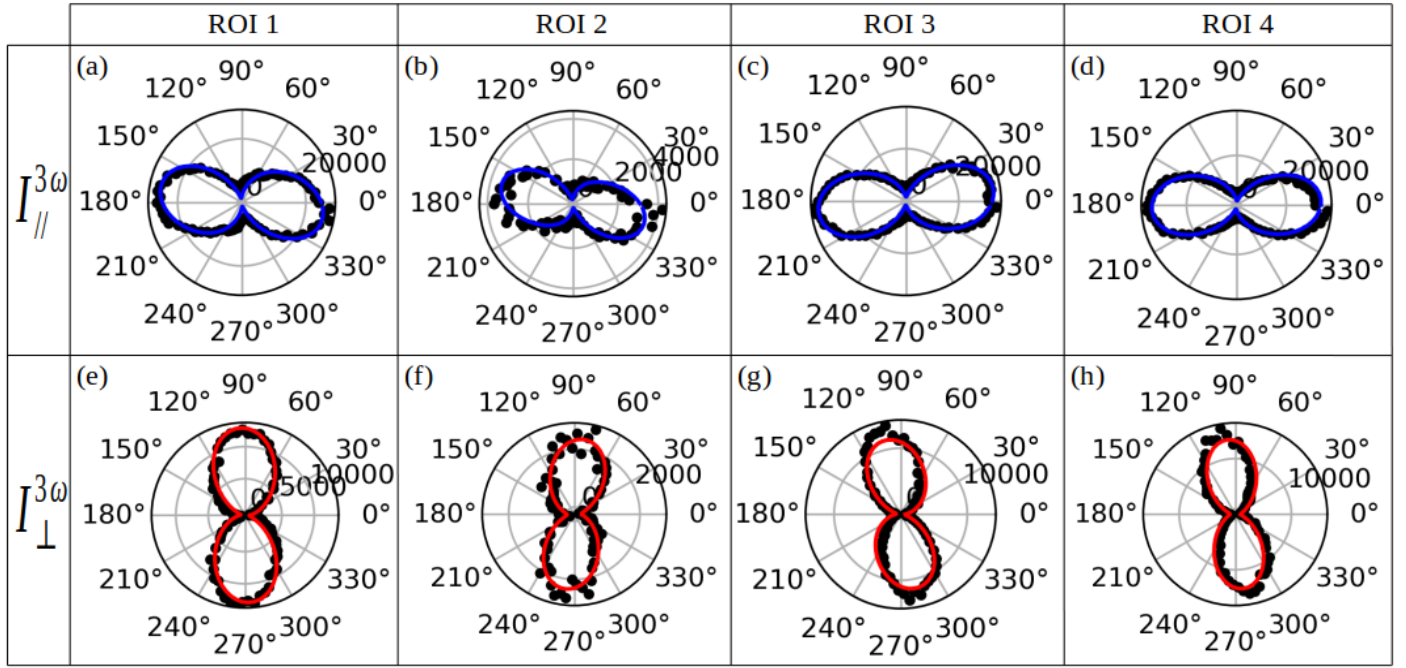

**Figure 5**. Experimental P-THG intensity modulations, $I_{\parallel}^{3\omega}$ and $I_{\perp}^{3\omega}$, as function of the linearly polarized excitation angle $\varphi$, presented in polar plots, for the ultrathin SnS crystals 1-4, corresponding to the ROIs marked in Figure 4. Black points represent the experimental data, while the blue and red lines represent the simultaneous fitting with Equation 3 ($I_{\parallel}^{3\omega}$) and 4 ($I_{\perp}^{3\omega}$), respectively. Using this fitting process, we are able to calculate the relative magnitudes of the $\chi^{(3)}$ tensor components, for each case, summarized in **Table 1**.

In Table 1, we summarize the fitting values for the relative magnitudes of the $\chi^{(3)}$ tensor components, for each SnS crystal under study. We also report the average values based on this sample of crystals, along with the standard deviation (std) values, as the errors of the average values. In order to obtain a graphical perspective of these average results, in Figure S10, Supporting Information, we plot them as Gaussian distributions, with the $\mu$ and $\sigma$ Gaussian parameters equal to the average and std values, respectively.

**Table 1.** Summary of the values of the relative magnitudes of the $\chi^{(3)}$ tensor components, $b$, $c$ and $d$, which we have experimentally calculated by simultaneously fitting both P-THG intensity components, $I_{\parallel}^{3\omega}$ and $I_{\perp}^{3\omega}$, with Equation 3 and 4, respectively. We present results for the ultrathin SnS crystals 1-6, corresponding to the ROIs marked in Figure 4 and S7, Supporting Information. The average values, along with the std values, as the errors of the average values, are also reported.

| Crystal | $b = \chi_{18}/\chi_{11}$ | $c = \chi_{22}/\chi_{11}$ | $d = \chi_{29}/\chi_{11}$ |
|---|---|---|---|
| 1 | 0.27 ± 0.02 | 0.99 ± 0.02 | 0.46 ± 0.02 |
| 2 | 0.40 ± 0.02 | 0.93 ± 0.03 | 0.49 ± 0.03 |
| 3 | 0.46 ± 0.02 | 1.00 ± 0.03 | 0.40 ± 0.02 |
| 4 | 0.21 ± 0.02 | 0.65 ± 0.05 | 0.16 ± 0.01 |
| 5 | 0.16 ± 0.02 | 0.74 ± 0.03 | 0.20 ± 0.01 |
| 6 | 0.09 ± 0.01 | 0.82 ± 0.02 | 0.40 ± 0.02 |
| Average | 0.26 ± 0.13 | 0.86 ± 0.13 | 0.35 ± 0.13 |

The results of Table 1 indicate the anisotropic nature of the $\chi^{(3)}$ tensor and the THG process in ultrathin SnS crystals. We conclude that the on-axis nonlinear susceptibility tensor elements, $\chi_{11}$ and $\chi_{22}$, are found to be considerably larger with respect to the off-axis elements, $\chi_{18}$ and $\chi_{29}$. The latter, expressed through *b* and *d*, are found to feature relatively similar magnitudes. These experimental results could be useful in future theoretical studies to evaluate or constrain theoretical models for THG in MXs.

It is noted that the values of the relative magnitudes of the $\chi^{(3)}$ tensor components exhibit variations among different crystals. Such variations have also been observed in other 2D materials, and have been attributed to deformation or defects in the crystal lattice, introduced during the exfoliation process.[35,40,41]

For reference, in Table S1, Supporting Information, we present a summary of literature reports, to our knowledge, with experimental results on the relative magnitudes of the $\chi^{(3)}$ tensor components, for other 2D materials. We observe that our conclusions are in qualitative agreement with most literature findings.

In **Figure 6**, we present both P-THG intensity components, together with the total P-THG intensity, $I^{3\omega} = I^{3\omega}_{\parallel} + I^{3\omega}_{\perp}$, for the ultrathin SnS crystal which corresponds to the red dashed square shown in Figure 4. In Figure S11, Supporting Information, we present the corresponding polar diagram for ROI 4 shown in Figure 4. The $I^{3\omega}_{\parallel}$ and $I^{3\omega}_{\perp}$ experimental data are simultaneously fitted with Equation 3 and 4, respectively, while the $I^{3\omega}$ data are fitted with Equation 6. The total P-THG intensity is found to be clearly anisotropic. Specifically, different angles $\varphi$ of the excitation linear polarization produce different THG intensities. Therefore, the above result confirms the in-plane anisotropic nature of the THG process in ultrathin SnS crystals. Furthermore, it offers the possibility of controlling the emitted THG intensity by tuning the angle $\varphi$.

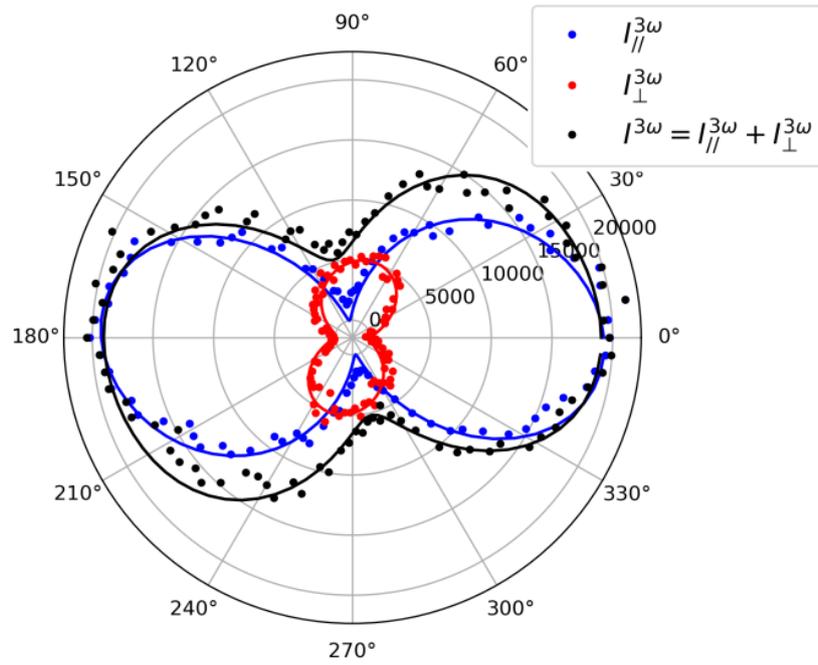

**Figure 6**. Experimental P-THG intensity modulations as function of the linearly polarized excitation angle $\varphi$, presented in a polar plot, for the ultrathin SnS crystal which corresponds to the red dashed square shown in Figure 4. Blue and red points represent $I_\parallel^{3\omega}$ and $I_\perp^{3\omega}$, respectively, while the black points represent the total P-THG intensity, $I^{3\omega} = I_\parallel^{3\omega} + I_\perp^{3\omega}$. The solid curves represent the respective theoretical fittings, with Equation 3, 4 and 6.

**2.4. THG Anisotropy Ratio**

Furthermore, through Equation 8, we have introduced a THG AR, which compares the total THG intensities for incident field polarization along the main crystallographic axes AC and ZZ. This AR has been correlated with the $\chi^{(3)}$ tensor ratio $c$, through Equation 8, and thus, can be directly calculated using the fitting results of $c$, reported in Table 1. The results are summarized in **Table 2**, for each ultrathin SnS crystal under study. The AR is found to be, on average, not equal to unity, suggesting that the total THG intensity is not equal when the excitation is along the AC direction compared to the case when it is along the ZZ direction. This AR allows to compare and classify different orthorhombic 2D materials, based on their degree of nonlinear optical anisotropy.

**Table 2.** Summary of the values we have experimentally calculated for the AR, through Equation 8. We present results for the ultrathin SnS crystals 1-6, corresponding to the ROIs marked in Figure 4 and S5, Supporting Information. The average value, along with the std value, as the error of the average value, are also reported.

| Crystal | Anisotropy ratio (AR) |
|---|---|
| 1 | 0.98 ± 0.04 |
| 2 | 0.86 ± 0.06 |
| 3 | 1.00 ± 0.06 |
| 4 | 0.42 ± 0.07 |
| 5 | 0.55 ± 0.04 |
| 6 | 0.67 ± 0.03 |
| Average | 0.75 ± 0.22 |

Finally, we address the effect of the thickness of a SnS crystal on its THG intensity. This can be realized by using a constructive interference model, where each layer contributes constructively in the detected THG signal.[30,50,59] In this context, in the case of a N-layer SnS crystal with zero twist-angle, the produced THG intensity is found to be analogous to $N^2$, where N is the number of layers (see Equations S6-S10, Supporting Information). This analysis allows to estimate the thickness of the SnS crystals (see Table S2 and the relevant discussion in section S8 in Supporting Information).

## 3. Conclusion

In conclusion, we have presented an all-optical, minimally invasive and rapid methodology, based on P-THG nonlinear optical imaging, to characterize the anisotropic properties of ultrathin SnS crystals, produced via LPE. We observe a variation of THG emission intensities for different polarization angles of the excitation field, demonstrating the in-plane anisotropic nature of the THG process. By using a polarizing beam splitter and two orthogonally placed detectors, we simultaneously record the intensity of the two orthogonal components of the THG field, while rotating the direction of the linear polarization of the excitation beam, enabling P-THG imaging. We then simultaneously fit the intensities of the two orthogonal components with a nonlinear optics model, which accounts for the orthorhombic crystal structure of 2D SnS. This approach enables the calculation of the relative magnitudes of the $\chi^{(3)}$ tensor components with high precision. Our results indicate that the on-axis nonlinear susceptibility tensor elements, $\chi_{11}$ and $\chi_{22}$, are considerably larger than the off-axis elements, $\chi_{18}$ and $\chi_{29}$, while the latter feature relatively similar magnitudes. We have also introduced and calculated a THG AR, which compares the total THG intensity upon excitation parallel to the AC direction compared to the case where it is parallel to the ZZ. Our results provide quantitative information on the effect of the in-plane anisotropy of ultrathin SnS on its nonlinear optical properties. We envisage that this work can introduce a useful method for studying the THG response of in-plane anisotropic 2D materials, towards fundamental studies, as well as the realization of polarization-sensitive nonlinear optical devices.

## 4. Experimental Section

*Nonlinear Optical Imaging Setup and Analysis*: The nonlinear optical imaging setup is based on a fs laser beam coupled to a microscope (Figure 2). The 1028 nm fs laser beam (FLINT FL1 Yb Oscillator, 1028 nm, ≈76 MHz, ≈36 fs, Light Conversion) passes through an optical parametric oscillator (APE Levante IR) which converts it to 1542 nm. The angle of the linear polarization of the excitation field is rotated, performing P-THG imaging, using an achromatic half-wave plate (AHWP10M-1600, Thorlabs), placed in a motorized rotation stage. The beam is guided into an inverted microscope (Axio Observer Z1, Carl Zeiss) by a pair of silver-coated galvanometric (galvo) mirrors (6215H, Cambridge Technology), allowing to raster-scan stationary samples. A pair of achromatic doublet lenses suitably expands the beam diameter to fill the back aperture of the objective lens (Plan-Apochromat 20×/0.8 NA, Carl Zeiss).

At the motorized turret box of the microscope, at 45° just below the objective lens, a silver mirror is used, which is insensitive to the polarization state of the laser beam. The objective lens tightly focuses the beam onto the sample, which, following light-matter interaction, produces THG radiation, which is collected in the forward detection geometry by a condenser lens (achromatic-aplanatic, 1.4 NA, Carl Zeiss). A polarizing beam splitter cube (CCM1-PBS251, Thorlabs) analyzes the THG signal into two orthogonal components, which are filtered by suitable short-pass (FF01-680/SP, Semrock) and narrow bandpass (FF01-514/3, Semrock) filters, placed just in front of the detectors, to cut off residual laser light and any other unwanted signal. The two orthogonally placed detectors, collecting the THG radiation, are based on photomultiplier tube modules (H9305-04, Hamamatsu). The diffraction-limited spatial resolution is ≈1.176 μm (0.61$\lambda_{exc}$/NA, with $\lambda_{exc}$ = 1542 nm, NA = 0.8).

The galvanometric mirrors and the photomultiplier tubes are connected to a connector block (BNC-2110, National Instruments Austin), which is interfaced to a PC through a DAQ (PCI 6259, National Instruments). The coordination of the detector recordings with the galvanometric mirrors for the image formation, as well as the movement of the motors, are carried out using LabView (National Instruments).

Each image presented here consists of 500×500 pixels. For the data analysis, the open-source Python programming language,[60] and the open-source ImageJ image analysis software,[61] are used. We note that in all experimental polar plots presented in this work: i) each ROI of pixels is treated as one pixel with intensity equal to the mean intensity in the ROI, and ii) we have selected a sample region without SnS crystals, whose intensity has been considered as noise and has been subtracted from our measurements. Regarding the adopted fitting procedure, the relative magnitudes of the $\chi^{(3)}$ tensor components are bounded in [0, 1]. Furthermore, experimentally, the throughput of the parallel and perpendicular P-THG intensities is different, and therefore, different scaling factors, α, were used during the fitting procedure.

*Sample Preparation and Characterization*: The LPE method is employed to isolate an ultrathin layer of SnS sheets;[30,43-47] see Supporting Information. The isolated ultrathin SnS sheets are characterized with UV-Vis spectroscopy, atomic force microscopy (AFM), Raman spectroscopy, and photoluminescence (PL) spectroscopy.

Figure S12a, Supporting Information, represents the UV-Vis extinction spectra of isolated SnS dispersion. The recorded extinction spectra of SnS exhibit a broad absorption profile in UV-Vis-NIR range, with a shoulder at 420 nm. The displayed extinction profile is consistent with the previously reported results.[43-46] This observation suggests the successful isolation of ultrathin layer of SnS sheets.

Moreover, AFM images are acquired to investigate the dimensionality of the isolated SnS sheets (Figure S12b, Supporting Information). The height profiles in the AFM images vary from 0.8 nm to 1.4 nm. The thickness of monolayer SnS is reported to be below 1 nm,[46] while the AFM measurements of liquid phase exfoliated SnS crystals have been reported to overestimate their thickness because of the solvent overlayer.[43] Therefore, our AFM measurements immensely suggest the presence of monolayer and bilayer crystals.

Furthermore, Raman spectroscopy is employed to quantify the layer number of the SnS crystals. Raman spectra of isolated SnS sheets reveal the expected optically active phonon modes;[45,46,61] Figure S13a, Supporting Information. Compared to the bulk crystal, the Raman peaks of the phonon modes of the ultrathin SnS sheets feature spectral blueshifts (Figure S13a, Supporting Information), which are consistent with our previous results of monolayer SnS sheets;[46] (see Supporting Information).

Finally, PL spectra of isolated thin layer SnS sheets are recorded, depicting a single broad emission peak, centered at ≈1.95 eV (Figure S13b, Supporting Information). The peak center of the PL emission closely matches the energy gap obtained from monolayer SnS.[46] All the above characterization techniques are further discussed in the Supporting Information.

**Supporting Information**
Supporting Information is available from the Wiley Online Library or from the author.

## Acknowledgments

E.S., G.K., L.M., S.P., and G.M.M. acknowledge financial support by the EU-funded DYNASTY project, ID: 101079179, under the Horizon Europe framework programme. E.S. and S.P. acknowledge funding by the Hellenic Foundation for Research and Innovation (H.F.R.I.) in the context of the project "Investigation of the protective roles of caloric restriction mimetics in myelin disruption via a novel, non-invasive advanced imaging approach_ID14772_MAYA". G.M.M. thanks Artem Larin (ITMO University) for a useful discussion on analyzing data related with nonlinear optical processes; Dionysios Xydias (IESL-FORTH) and Maria Kefalogianni (IESL-FORTH) for their contribution while sharing the nonlinear optical setup; and Matina Vlahou (IESL-FORTH) for designing the illustration of the nonlinear optical setup.

## Author Contributions

S.P., L.M., E.S., and G.K. guided the research. G.M.M. and S.P. conducted the nonlinear optical experiments. L.M., G.M.M., and S.P. derived the theoretical equations. G.M.M., S.P., L.M., and A.L. conducted the data analysis. A.S.S. prepared and characterized the samples. A.L. provided technical support. G.M.M., L.M., S.P., E.S., and A.S.S. prepared the manuscript.

## Conflict of Interest

The authors declare no conflict of interest.

## Data Availability Statement

The data that support the findings of this study are available from the corresponding author upon reasonable request.

# Supporting Information

## Anisotropic Third Harmonic Generation in Two-Dimensional Tin Sulfide

*George Miltos Maragkakis\*, Sotiris Psilodimitrakopoulos\*, Leonidas Mouchliadis, Abdus Salam Sarkar, Andreas Lemonis, George Kioseoglou, and Emmanuel Stratakis\**

G. M. Maragkakis, S. Psilodimitrakopoulos, L. Mouchliadis, A. S. Sarkar, A. Lemonis, G. Kioseoglou, E. Stratakis

Institute of Electronic Structure and Laser Foundation for Research and Technology-Hellas, Heraklion, Crete, 71110, Greece

E-mails: gmaragkakis@physics.uoc.gr; sopsilo@iesl.forth.gr; stratak@iesl.forth.gr

G. M. Maragkakis, E. Stratakis

Department of Physics, University of Crete, Heraklion, Crete, 71003, Greece

G. Kioseoglou

Department of Materials Science and Technology, University of Crete, Heraklion, Crete, 71003, Greece

## S1. Theoretical formulation of P-THG in 2D SnS

The THG equation in two-dimensional (2D) SnS is expressed in crystal coordinates as in Equation 2, main manuscript.

We then transform Equation 2, main manuscript, back to laboratory coordinates by multiplying with the rotation matrix $\begin{pmatrix} \cos\theta & -\sin\theta \\ \sin\theta & \cos\theta \end{pmatrix}$, obtaining:

$$\begin{pmatrix} P_X^{3\omega} \\ P_Y^{3\omega} \end{pmatrix} \sim \begin{pmatrix} (\chi_{11}\cos^2(\varphi-\theta) + 3\chi_{18}\sin^2(\varphi-\theta))\cos\theta\cos(\varphi-\theta) - (\chi_{22}\sin^2(\varphi-\theta) + 3\chi_{29}\cos^2(\varphi-\theta))\sin\theta\sin(\varphi-\theta) \\ (\chi_{11}\cos^2(\varphi-\theta) + 3\chi_{18}\sin^2(\varphi-\theta))\sin\theta\cos(\varphi-\theta) + (\chi_{22}\sin^2(\varphi-\theta) + 3\chi_{29}\cos^2(\varphi-\theta))\cos\theta\sin(\varphi-\theta) \end{pmatrix}$$
(S1)

By using a polarizing beam splitter in front of the detectors, we collect the two orthogonal components of the THG field. These components, $P_\parallel^{3\omega} = P_X^{3\omega}$ and $P_\perp^{3\omega} = P_Y^{3\omega}$, can be described by multiplying with the Jones matrix $\begin{pmatrix} \cos^2\zeta & \sin\zeta\cos\zeta \\ \sin\zeta\cos\zeta & \sin^2\zeta \end{pmatrix}$ of a linear polarizer. Here, $\zeta$ is the angle of the transmission axis of the polarizer with respect to $X$-axis, and we set $\zeta = 0°$ for $P_\parallel^{3\omega}$, and $\zeta = 90°$ for $P_\perp^{3\omega}$, i.e., axis of transmission parallel to $X$ and $Y$-axis, respectively. Then, the intensities of these two orthogonal THG components, $I_\parallel^{3\omega} = |P_\parallel^{3\omega}|^2$ and $I_\perp^{3\omega} = |P_\perp^{3\omega}|^2$, are calculated to be:

$$I_\parallel^{3\omega} \sim [(\chi_{11}\cos^2(\varphi-\theta) + 3\chi_{18}\sin^2(\varphi-\theta))\cos\theta\cos(\varphi-\theta) - (\chi_{22}\sin^2(\varphi-\theta) + 3\chi_{29}\cos^2(\varphi-\theta))\sin\theta\sin(\varphi-\theta)]^2 \quad (S2)$$

$$I_\perp^{3\omega} \sim [(\chi_{11}\cos^2(\varphi-\theta) + 3\chi_{18}\sin^2(\varphi-\theta))\sin\theta\cos(\varphi-\theta) + (\chi_{22}\sin^2(\varphi-\theta) + 3\chi_{29}\cos^2(\varphi-\theta))\cos\theta\sin(\varphi-\theta)]^2 \quad (S3)$$

These relationships are expressed in terms of the absolute values of the $\chi^{(3)}$ tensor elements. Instead, they can be expressed in terms of dimensionless ratios of the $\chi^{(3)}$ tensor elements, obtaining Equation 3 and 4, main manuscript.

We also note that the produced THG field is also linearly polarized at an angle $\theta_{THG}$ given by:

$$\theta_{THG} = \tan^{-1}\left\{\frac{\chi_{22}\tan^2(\varphi-\theta) + 3\chi_{29}}{\chi_{11} + 3\chi_{18}\tan^2(\varphi-\theta)}\tan(\varphi-\theta)\right\} \quad (S4)$$

which for the special case of $\varphi = \theta + 45°$ yields:

$$\theta_{THG} = \tan^{-1}\left\{\frac{\chi_{22} + 3\chi_{29}}{\chi_{11} + 3\chi_{18}}\right\} \quad (S5)$$

## S2. Numerical simulations on the effect of different values of the parameters on the P-THG intensity components in MXs

### S2.1. Varying the ratios of the $\chi^{(3)}$ tensor components

In the following figures, we present numerical simulations on the effect of different values of the relative magnitudes of the $\chi^{(3)}$ tensor components, $b$ (**Figure S1**), $c$ (**Figure S2**) and $d$ (**Figure S3**), defined by Equation 5, on the polarization-resolved third harmonic generation (P-THG) intensity components produced by a group IV monochalcogenide (MX), described by Equation 3 and 4. In each case, we vary only one ratio of the $\chi^{(3)}$ tensor components as reported in each figure, while keeping the rest ratios fixed to the average values we have eventually, experimentally calculated from our analysis, reported in Table 1. In all cases, the armchair (AC) direction $\theta$ is kept fixed to 50°.

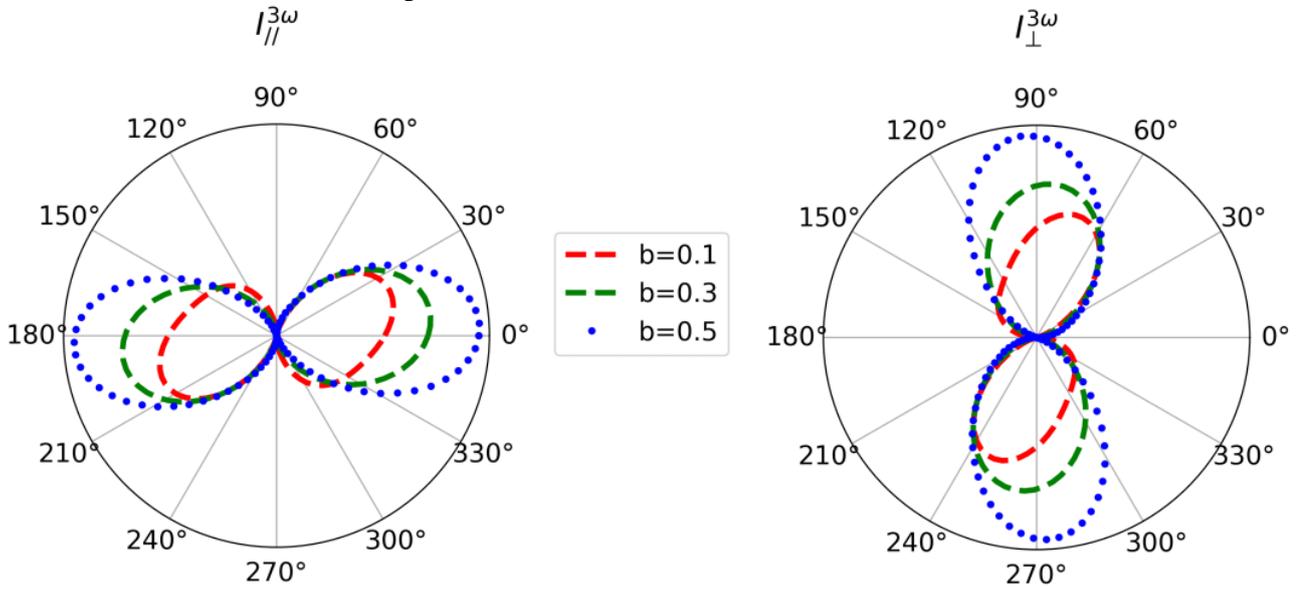

**Figure S1**. Numerical simulations on the effect of different values of the $\chi^{(3)}$ tensor ratio $b$ (defined by Equation 5) on the P-THG intensity components produced by a MX, described by Equation 3 and 4. We plot $I_{\parallel}^{3\omega}$ (left) and $I_{\perp}^{3\omega}$ (right), in polar diagrams, as function of the orientation of the linearly polarized excitation angle $\varphi$.

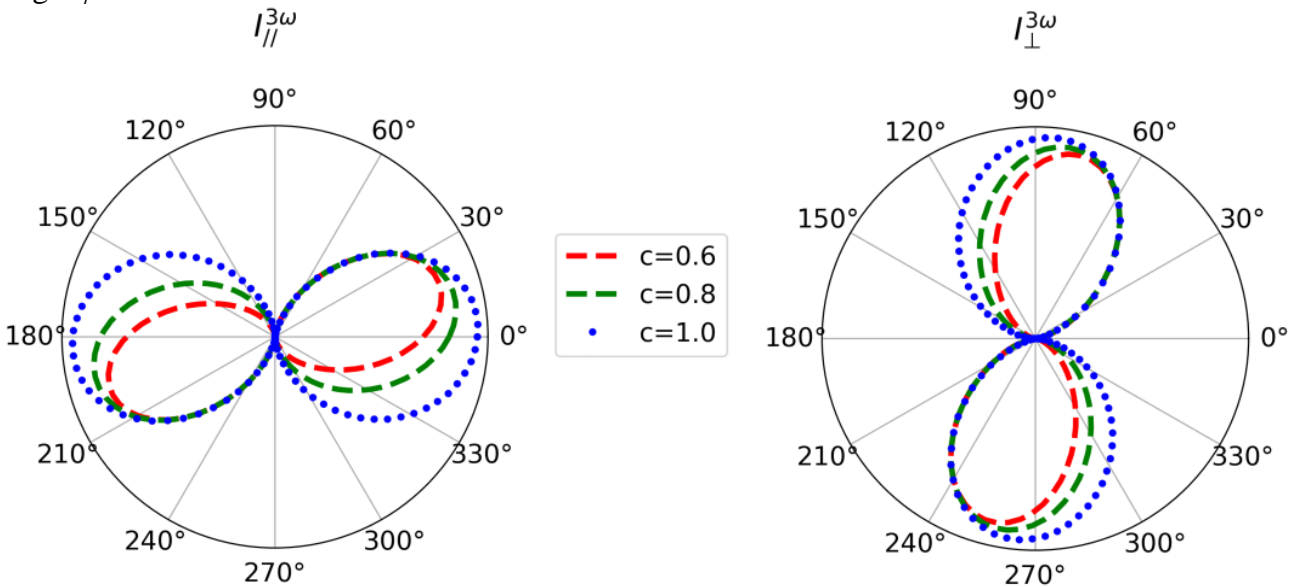

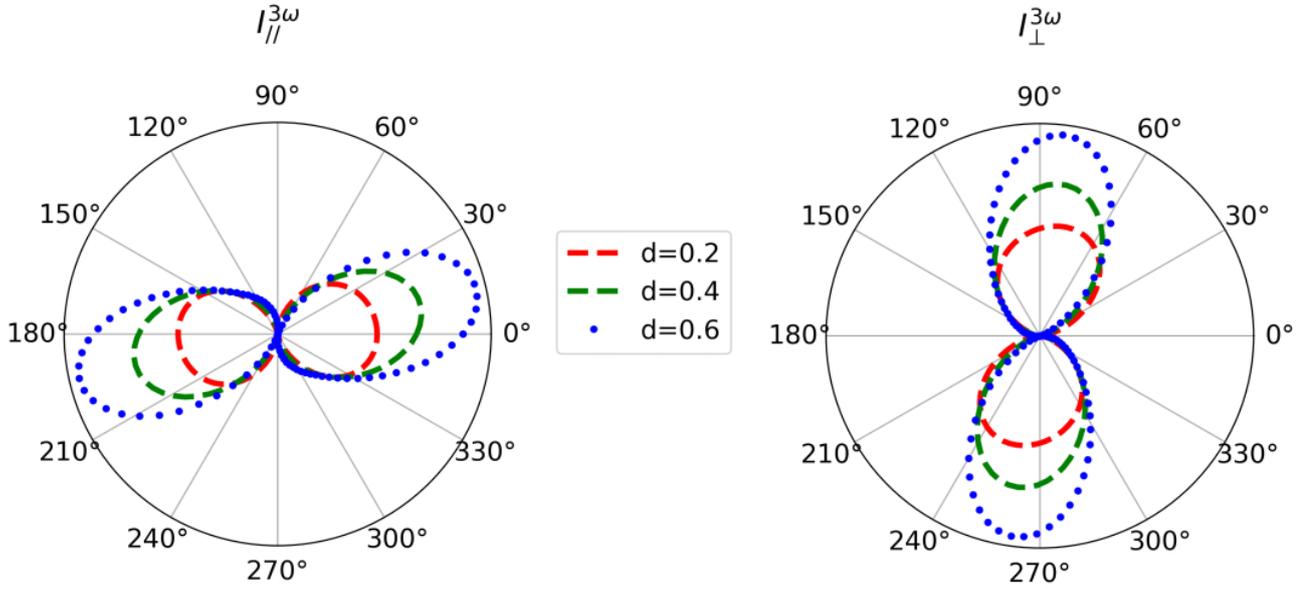

**Figure S2.** Numerical simulations on the effect of different values of the $\chi^{(3)}$ tensor ratio $c$ (defined by Equation 5) on the P-THG intensity components produced by a MX, described by Equation 3 and 4. We plot $I_\parallel^{3\omega}$ (left) and $I_\perp^{3\omega}$ (right), in polar diagrams, as function of the orientation of the linearly polarized excitation angle $\varphi$.

**Figure S3.** Numerical simulations on the effect of different values of the $\chi^{(3)}$ tensor ratio $d$ (defined by Equation 5) on the P-THG intensity components produced by a MX, described by Equation 3 and 4. We plot $I_\parallel^{3\omega}$ (left) and $I_\perp^{3\omega}$ (right), in polar diagrams, as function of the orientation of the linearly polarized excitation angle $\varphi$.

### S2.2. Varying the armchair angle

In **Figure S4**, we present numerical simulations on the effect of different values of the AC angle $\theta$ on the P-THG intensity components produced by a MX, described by Equation 3 and 4. The relative magnitudes of the $\chi^{(3)}$ tensor components, $b$, $c$ and $d$ are kept fixed to the average values we have eventually, experimentally calculated from our analysis, reported in Table 1.

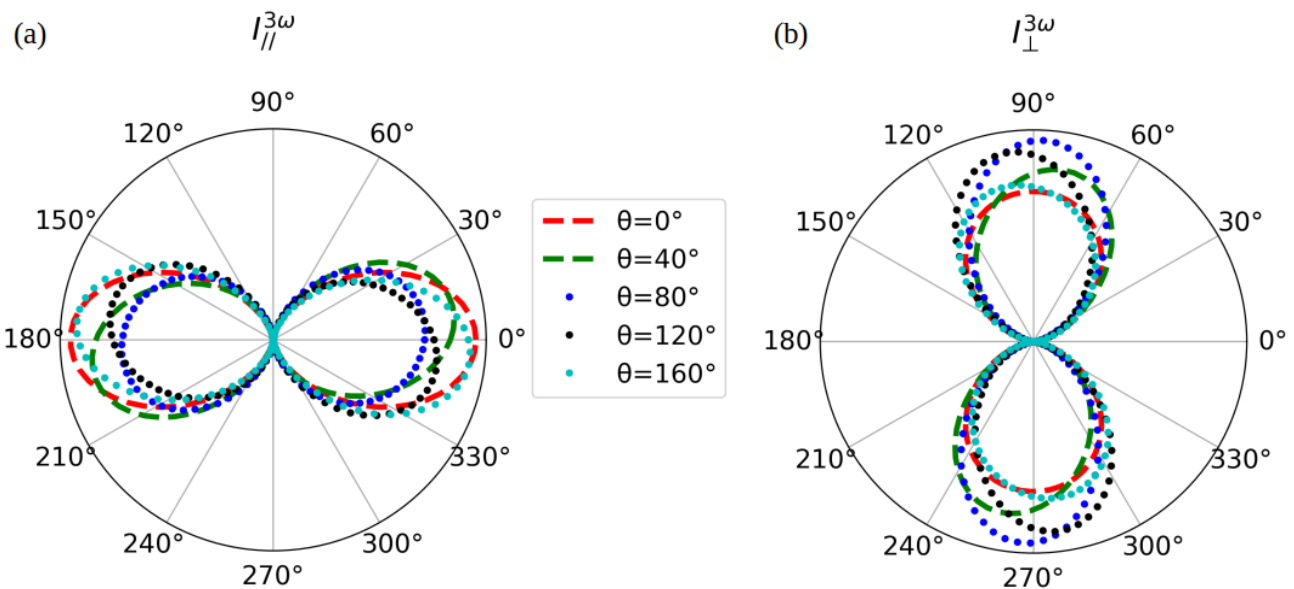

**Figure S4.** Numerical simulations on the effect of different values of the AC angle $\theta$ on the P-THG intensity components produced by a MX, described by Equation 3 and 4. We plot a) $I_\parallel^{3\omega}$ and b) $I_\perp^{3\omega}$, in polar diagrams, as function of the orientation of the linearly polarized excitation angle $\varphi$.

## S3. Estimation of the smallest detectable change in the armchair angle

Here, we aim to estimate the smallest detectable change in the AC crystallographic direction of a 2D SnS crystal, based on the P-THG measurements. For this purpose, we have performed numerical simulations using "ideal" P-THG data, subject only to Poisson noise, which is commonly used to describe the effect of photon noise in our detectors, namely photomultiplier tubes.[1] In particular, we have generated ninety P-THG values, for both $I_{\parallel}^{3\omega}$ and $I_{\perp}^{3\omega}$ through Equations 3 and 4, respectively, for $\varphi \in [0°, 360°)$ with step of 4°, using fixed values of the parameters of the model, namely: the AC angle, the scaling factors α ($a_{\parallel} = 2$ and $a_{\perp} = 1$), and the relative magnitudes of the $\chi^{(3)}$ tensor components b, c and d (equal to the average values we have eventually, experimentally calculated from our analysis, reported in Table 1). In these "ideal" P-THG values, we have applied 1000 realizations of Poisson noise, and then performed the same simultaneous fitting procedure adopted to fit the real experimental data, for each realization. However, in this case, only the parameter θ is fitted, and not α, b, c and d. In this way, we have extracted the angle θ, for each realization of Poisson noise. We have repeated this analysis for various angles θ. An example is presented in **Figure S5**, for θ = 50°.

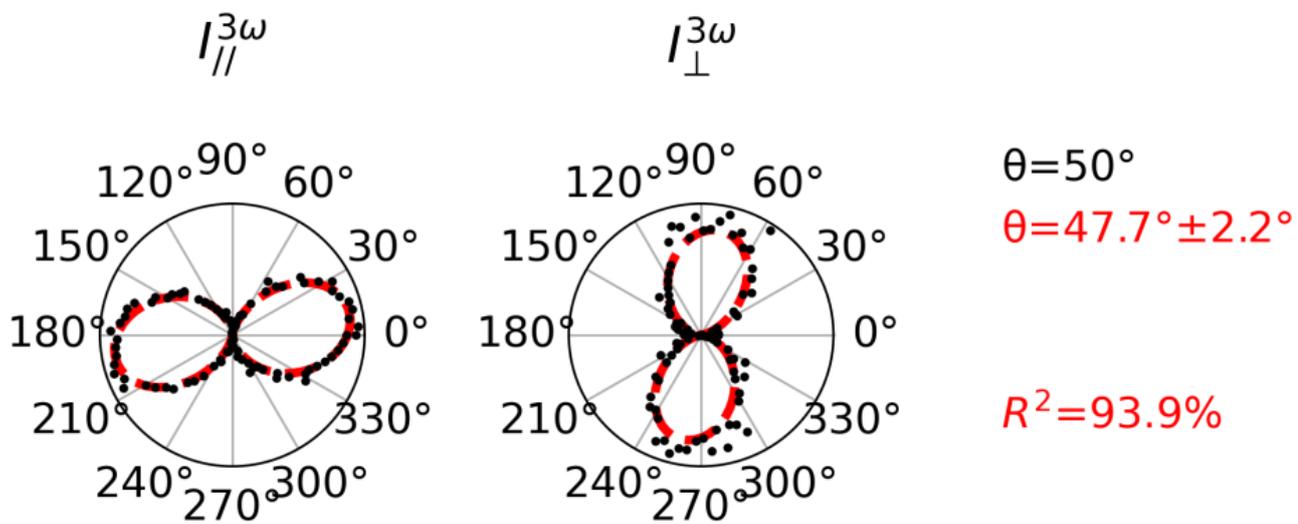

**Figure S5.** Numerical simulation which investigates the precision with which our theoretical model and fitting process fit the data and calculate the AC angle θ. In order to perform this evaluation, we have generated "ideal" P-THG values, for both $I_{\parallel}^{3\omega}$ and $I_{\perp}^{3\omega}$ (denoted by black dots in the polar plots), using fixed values of the parameters of the model (θ = 50° in this example). In these values, we have applied 1000 realizations of Poisson noise, and performed the same simultaneous fitting procedure adopted to fit the real experimental data, for each realization; however, in this case, only the parameter θ is fitted. In this way, we have extracted the angle θ, reported on the right in red color, calculated as the average value that resulted from the set of the 1000 realizations of Poisson noise; as well the respective fitting curve (denoted by red color in the polar plots). The average quality of fitting $R^2$ is also reported.

In the context we have described, as the smallest detectable change in the AC angle, we have utilized the standard deviation value, $\sigma_\theta$, that resulted from the 1000 realizations of Poisson noise. For "ideal" data (quality of fitting $R^2 = 100\%$), $\sigma_\theta$ is found to be zero (blue points in **Figure S6**). Deviation from the "ideal" data, manifested through the Poisson noise, provides error in the calculation of the angle θ. Interestingly, $\sigma_\theta$ is found to depend on the angle θ. It is estimated to be in the range $1.5° < \sigma_\theta \leq 2.3°$ for quality of fitting $R^2 > 92\%$, and in the range $2.1° < \sigma_\theta \leq 3.4°$ for quality of fitting $R^2 > 85\%$ (green and red points, respectively, in Figure S6).

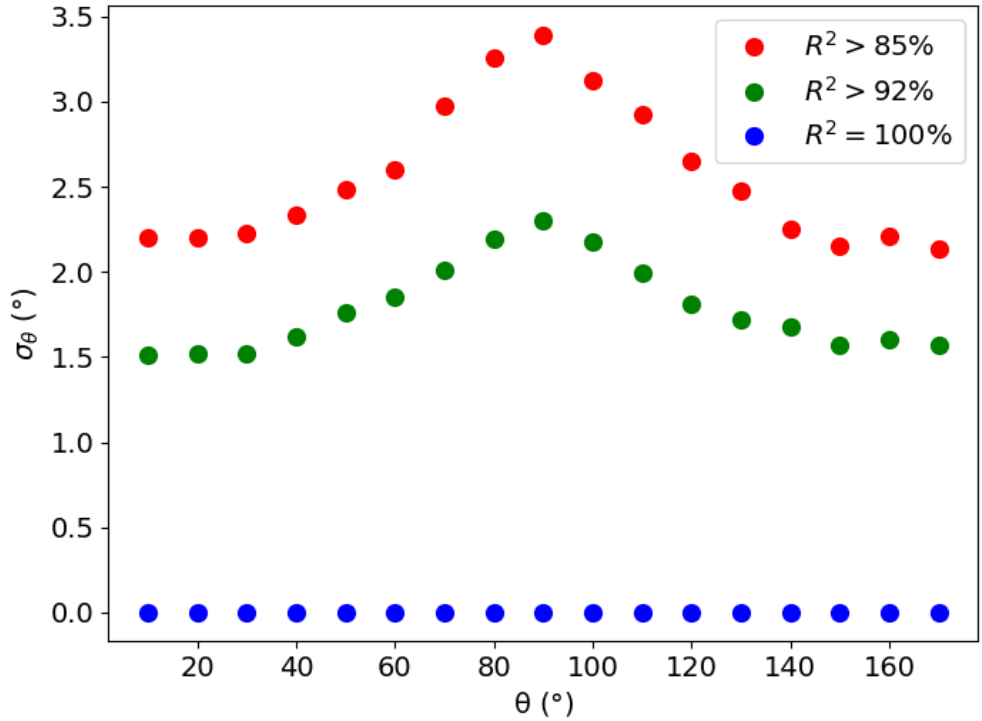

**Figure S6.** Numerical simulation which investigates the smallest detectable change in the AC angle θ of a 2D SnS crystal, based on the P-THG measurements. In order to perform this evaluation, we have generated "ideal" P-THG values, for both $I_\parallel^{3\omega}$ and $I_\perp^{3\omega}$, using fixed values of the parameters of the model. In these values, we have applied 1000 realizations of Poisson noise, and performed the same simultaneous fitting procedure adopted to fit the real experimental data. However, in this case, only the parameter θ is fitted. In this way, we have extracted the angle θ, for each realization of Poisson noise. Here we plot the standard deviation value, $\sigma_\theta$, that resulted from the 1000 realizations of Poisson noise, as function of the values of the angles θ that were used to generate the simulated data. Results for three different quality of fittings $R^2$ are reported.

**S4. Experimental P-THG imaging**

In **Figure S7**, we present the sum of all ninety collected P-THG intensity images of $I_\parallel^{3\omega}$, for all orientations $\varphi$ of the excitation linear polarization, corresponding to a field of view containing ultrathin tin (II) sulfide (SnS) crystals. The two crystals we have focused in our analysis are marked.

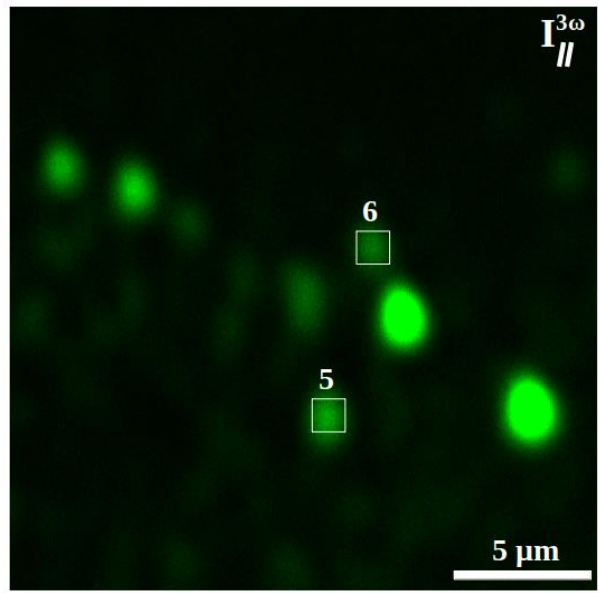

**Figure S7.** Sum of all ninety collected P-THG intensity images of $I_\parallel^{3\omega}$, for all orientations $\varphi$ of the excitation linear polarization, corresponding to a field of view containing ultrathin SnS crystals. The two crystals we have focused in our analysis are marked. Brighter color indicates higher THG intensity.

## S5. Experimental fitting analysis

In **Figure S8**, we experimentally demonstrate that fitting only one P-THG intensity component is not sufficient to obtain reliable fitting results. An example is presented, for the crystal shown with the red square in the THG images (left), which corresponds to crystal 3 shown in Figure 4. The white square in the same images shows a region without SnS crystals, whose intensity has been considered as noise and has been subtracted from our measurements, following the methodology adopted in the main manuscript.

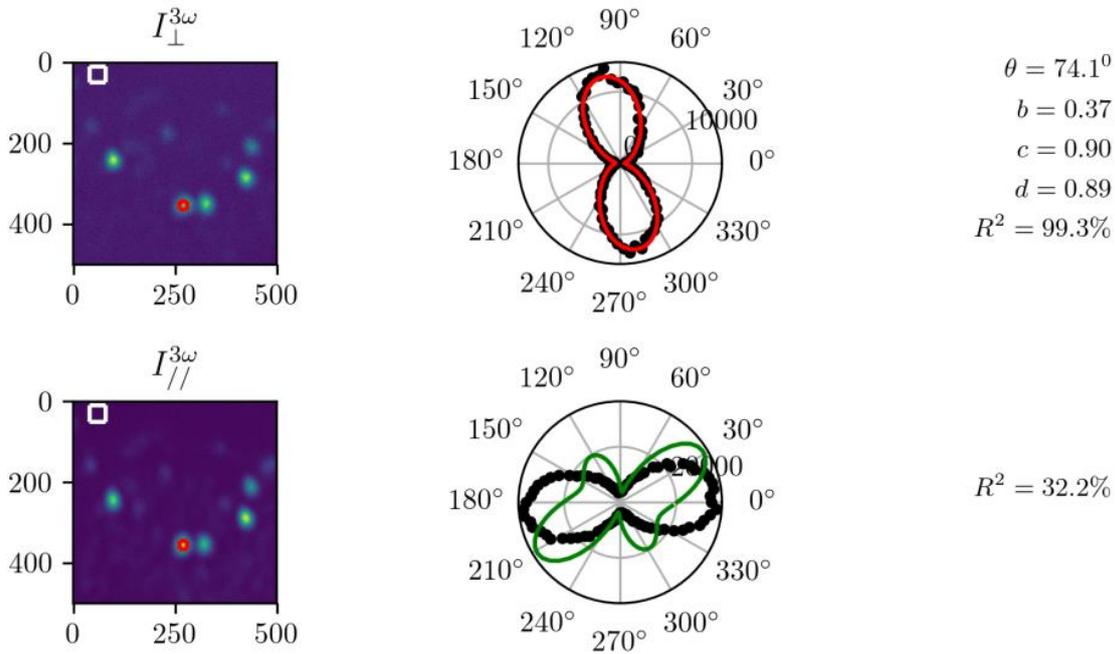

**Figure S8.** Example showing experimentally that fitting only one P-THG intensity component (either $I_\parallel^{3\omega}$ or $I_\perp^{3\omega}$) is not sufficient to obtain reliable fitting results. On the left, we show THG images from the same field of view, where the THG intensity is integrated for all orientations $\varphi$ of the excitation linear polarization. In the polar plots (center), black points represent the experimental P-THG data, for the SnS crystal shown with the red square. Here we have fitted $I_\perp^{3\omega}$ with Equation 4 (red curve), and then we have tested the fitting parameter values (reported on the right) on $I_\parallel^{3\omega}$, through Equation 3 (green curve). Although this set of fitting parameter values fits very well the one component (quality of fitting 99.3%), it does not fit the other (quality of fitting 32.2%), producing a four-lobe pattern instead of a two-lobe one.

In **Figure S9**, we present the simultaneous fitting to both P-THG intensity components, for the crystals 5 and 6, corresponding to the regions of interest (ROIs) shown in Figure S7.

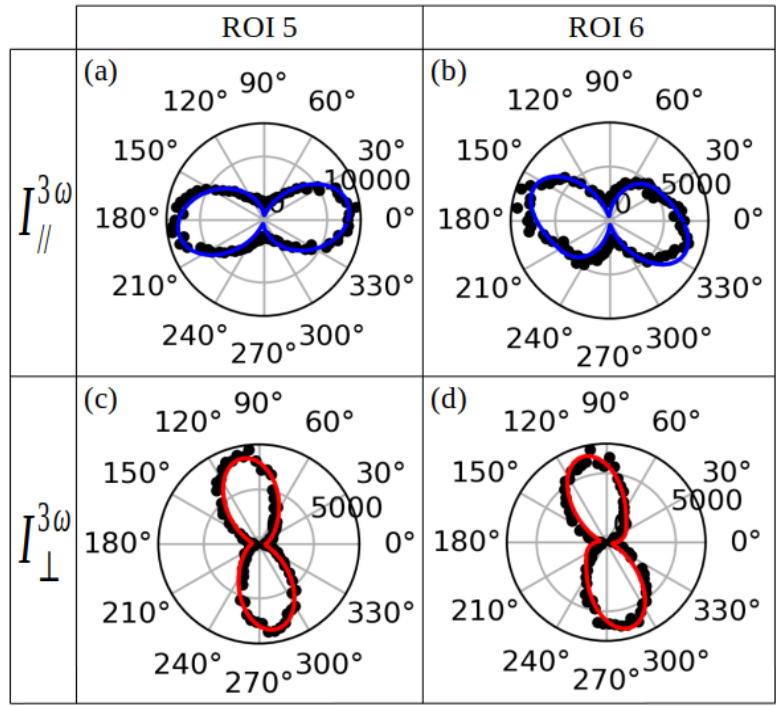

**Figure S9.** Experimental P-THG intensity modulations, $I_{\parallel}^{3\omega}$ and $I_{\perp}^{3\omega}$, as function of the linearly polarized excitation angle $\varphi$, presented in polar plots, for the ultrathin SnS crystals 5 and 6, corresponding to the ROIs marked in Figure S7. Black points represent the experimental data, while the blue and red lines represent the simultaneous fitting with Equation 3 ($I_{\parallel}^{3\omega}$) and Equation 4 ($I_{\perp}^{3\omega}$), respectively. With this fitting, we are able to calculate the relative magnitudes of the $\chi^{(3)}$ tensor components, $b$, $c$ and $d$, for each case, summarized in Table 1.

## S6. Relative magnitudes of the $\chi^{(3)}$ tensor components

In Table 1, we have reported the average values of the relative magnitudes of the $\chi^{(3)}$ tensor components, $b$, $c$ and $d$, calculated from the fitting results on the sample of the six ultrathin SnS crystals under study. In **Figure S10**, we plot these average values as Gaussian distributions, with the $\mu$ and $\sigma$ Gaussian parameters equal to the average and standard deviation (std) values, respectively.

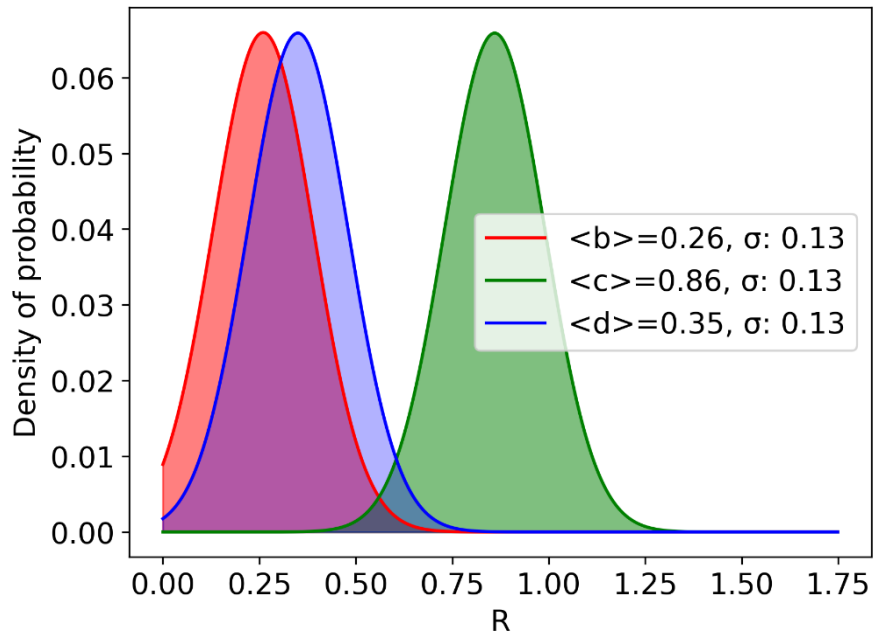

**Figure S10.** Graphical representation of the average values we have experimentally calculated for the relative magnitudes of the $\chi^{(3)}$ tensor components, *b*, *c* and *d*, which are here denoted by *R*. These average values, reported in Table 1, are plotted here as Gaussian distributions, with the *μ* and *σ* Gaussian parameters equal to the average and std values, respectively.

In **Table S1**, we summarize literature findings, to our knowledge, which report experimental results on the relative magnitudes of the $\chi^{(3)}$ tensor components, for other 2D materials.

**Table S1.** Summary of literature reports, to our knowledge, with experimental results on the relative magnitudes of the $\chi^{(3)}$ tensor components, for other 2D materials [2-7]. (*) For GeSe and GeAs, the values we present here are the average values we have calculated based on a sample of two (ref. [2]) and seven (ref. [6]) crystals, respectively, which are reported in the corresponding studies (along with the std values as the errors of the average values).

| Crystal | Crystal structure | $b = \chi_{18}/\chi_{11}$ | $c = \chi_{22}/\chi_{11}$ | $d = \chi_{29}/\chi_{11}$ |
|---|---|---|---|---|
| SnS (This work) | Orthorhombic | 0.26 ± 0.13 | 0.86 ± 0.13 | 0.35 ± 0.13 |
| GeSe (ref. [2]) (*) | Orthorhombic | 0.26 ± 0.01 | 0.54 ± 0.06 | 0.4 ± 0.1 |
| Black phosphorus (ref. [3]) | Orthorhombic | 0.46 | 0.5 | 0.35 |
| Black phosphorus (ref. [4]) | Orthorhombic | Almost zero | Comparable with d | Comparable with c |
| SiP (ref. [5]) | Orthorhombic | - | 0.72 ± 0.08 | - |
| GeAs (ref. [6]) (*) | Monoclinic | 0.21 ± 0.04 | 0.48 ± 0.04 | 0.19 ± 0.01 |
| $As_2S_3$ (ref. [7]) | Monoclinic | 0.29 | 0.60 | 0.29 |

**S7. Total P-THG intensity**

In **Figure S11**, we present both P-THG intensity components, together with the total P-THG intensity, $I^{3\omega} = I^{3\omega}_{\parallel} + I^{3\omega}_{\perp}$, for the ultrathin SnS crystal which corresponds to ROI 4 shown in Figure 4. The $I^{3\omega}_{\parallel}$ and $I^{3\omega}_{\perp}$ experimental data are simultaneously fitted with Equation 3 and 4, respectively, while the $I^{3\omega}$ data are fitted with Equation 6.

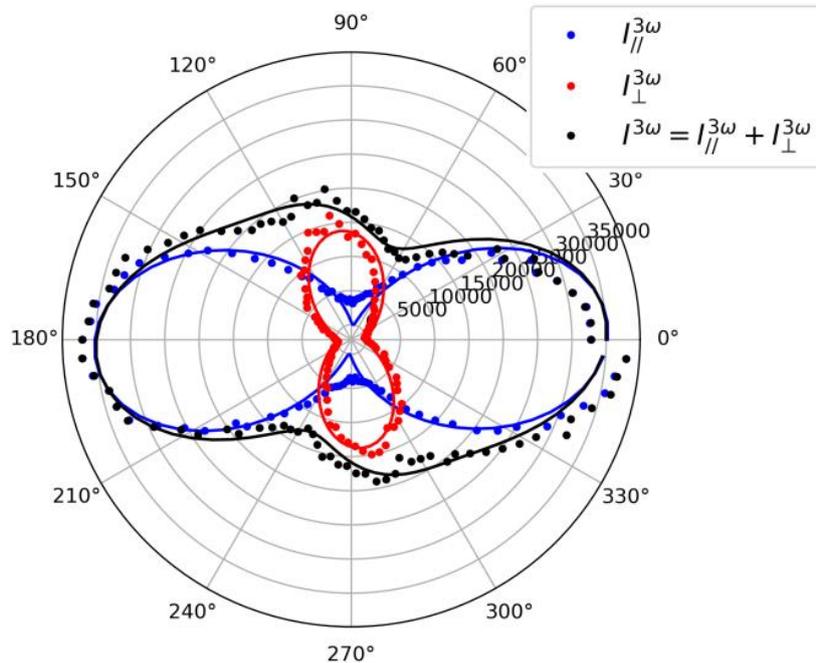

**Figure S11**. Experimental P-THG intensity modulations as function of the linearly polarized excitation angle $\varphi$, presented in a polar plot, for the ultrathin SnS crystal which corresponds to ROI 4 shown in Figure 4. Blue and red points represent $I_\parallel^{3\omega}$ and $I_\perp^{3\omega}$, respectively, while the black points represent the total P-THG intensity, $I^{3\omega} = I_\parallel^{3\omega} + I_\perp^{3\omega}$. The solid curves represent the respective theoretical fittings, with Equation 3, 4 and 6.

**S8. THG from N-layer SnS crystals**

In order to address the thickness-dependent THG intensity generated from a layered group IV MX crystal, with N number of layers, we can use a constructive interference model.[8-10]. The third harmonic field arising will have the form of vector superposition:

$$\boldsymbol{E}^{3\omega} = \boldsymbol{E}_1^{3\omega} + \boldsymbol{E}_2^{3\omega} + \ldots + \boldsymbol{E}_N^{3\omega} \quad (S6)$$

where the indices denote the third harmonic contribution from the corresponding layers. The total THG intensity produced by the N-layer structure, will then be:

$$I^{3\omega} = |\boldsymbol{E}_1|^2 + |\boldsymbol{E}_2|^2 + \ldots + |\boldsymbol{E}_N|^2 + 2\boldsymbol{E}_1 \cdot \boldsymbol{E}_2 + \ldots + 2\boldsymbol{E}_{N-1} \cdot \boldsymbol{E}_N \quad (S7)$$

$$I^{3\omega} = I_1 + I_2 + \ldots + I_N + 2\sqrt{I_1 I_2} \cos\delta_{1,2} + \ldots + 2\sqrt{I_{N-1} I_N} \cos\delta_{N-1,N} \quad (S8)$$

where $\delta_{i,j}$, $i,j = 1,2,\ldots,N$ denote the relative angle between layers $i$ and $j$, i.e., the twist-angles, and the frequency index $3\omega$ is suppressed for simplicity. If we assume for simplicity that the THG intensity from the individual layers is equal ($I_1 = I_2 = \ldots = I_N = I_{ML}^{3\omega}$), and that the layers are aligned (i.e., all twist-angles are zero), we find that:

$$I^{3\omega} = N I_{ML}^{3\omega} + I_{ML}^{3\omega} N(N-1) \quad (S9)$$

$$I^{3\omega} = N^2 I_{ML}^{3\omega} \quad (S10)$$

Therefore, we find that the THG intensity from 2D crystals with zero twist-angle is analogous to $N^2$, where N is the number of layers. It is valid for 2D crystals, where each layer contributes constructively in the detected THG signal. It has been experimentally verified for $WSe_2$ crystals.[11]

The above analysis allows to interpret the reason behind the considerably higher THG intensity exhibited by some SnS crystals compared to others, as illustrated in Figure 4a, 4b and S7. We note that the integrated THG intensity from each SnS crystal shown in these figures (i.e., sum of all ninety THG intensities acquired for all the different excitation polarization angles $\varphi \in [0°, 360°)$ with step of 4°) is not polarization-dependent anymore, and thus differences in the THG intensities among the different crystals in the same image are attributed to differences in their number of layers. In order to quantify these differences, **Table S2** summarizes the maximum THG intensities detected in each ROI. We characterize as monolayer the crystal with the minimum intensity, i.e., ROI 2. Then, the number of layers $N$ for the other crystals can be estimated using Equation S10 ($N = \sqrt{I^{3\omega}/I_{ML}^{3\omega}}$), with the results summarized in Table S2. We find that, according to the above constructive interference model, the $N^2$ times THG signal dependency, which appeared in our experimental data, could occur from bilayer and trilayer SnS crystals.

Interestingly, in these ROIs, we observe a variation in the THG intensity and consequently in the calculated number of layers (i.e., we find not exactly $N = 2$ and $N = 3$ in Table S2, but rather $N = 2 \pm 0.2$ and $N = 3 \pm 0.2$). This variation may be attributed to solvent overlayer, which could change the z-positions of the SnS crystals and place them slightly out of focus.[10] This solvent residual is known to affect the measurement of the thickness of liquid phase exfoliated SnS flakes.[10,12]

**Table S2**. Summary of the maximum THG intensity detected from the ultrathin SnS crystals 1-6, corresponding to the ROIs marked in Figure 4 and S7. As monolayer is characterized the crystal with the minimum intensity, i.e., ROI 2, while the number of layers *N* for the other crystals is estimated using Equation S10.

| Crystal | Max. THG intensity (x $10^5$) (arb. units) | Number of layers N |
|---|---|---|
| 1 | 16.4 | 2.9 |
| 2 | 1.9 | 1 |
| 3 | 17.9 | 3.1 |
| 4 | 19.8 | 3.2 |
| 5 | 9.4 | 2.2 |
| 6 | 6.5 | 1.8 |

## S9. Material preparation and characterization

### S9.1. Preparation of SnS sheets

In order to isolate an ultrathin layer of SnS sheets, the liquid phase exfoliation method is employed.[10,12-16] In particular, tin (II) sulfide granular (>99.99%, Lot No-40105, Sigma Aldrich) trace metal basis (3mg/mL) is dissolved in acetone (≥99.5%, Honeywell), and the glass bottle is sealed under molecular nitrogen ($N_2$) atmosphere. The SnS solution is ultrasonicated (Elma S 30 H, Elma Schmidbauer GmbH), using a bath sonicator with power 80 W and frequency 37 kHz, for 20 h. The ultrasonicated solution is centrifuged at 8000 rpm for 15 min. The suspended solution containing SnS is collected and used for further experimental investigations.

### S9.2. Material characterization

The isolated ultrathin SnS sheets are characterized with UV-Vis spectroscopy (**Figure S12**a), atomic force microscopy (AFM) (Figure S12b), Raman spectroscopy (**Figure S13**a), and photoluminescence (PL) spectroscopy (Figure S13b). The results presented in Figure S12 and S13 are discussed in the main manuscript.

### S9.3. Raman spectroscopy

The layer number of SnS thin sheets is quantified using Raman spectroscopy. Figure S13a represents the Raman spectra of bulk SnS and the thin layer of SnS sheets. The spectra depict several vibrational modes. In particular, the SnS crystal belongs to the orthorhombic point group $C_{2v}$ (mm2), with 24 phonon modes at the center of the Brillouin zone, $\Gamma$, expressed as:[14,17]

$$\Gamma = 4A_g + 2B_{1g} + 4B_{2g} + 2B_{3g} + 2A_u + 4B_{1u} + 2B_{2u} + 4B_{3u} \quad (S11)$$

where $A_g$, $B_{1g}$, $B_{2g}$ and $B_{3g}$ are optically active Raman modes. The room temperature Raman spectra of our isolated SnS sheets feature the expected optically active phonon modes (Figure S13a). The phonon modes peak at ≈100, ≈164 and ≈194 cm$^{-1}$, and are associated with $A_g(1)$, $B_{3g}$ and $A_g(2)$ modes, respectively. The isolated thin layer of SnS sheets reveal the prominent Raman peaks at ≈167 and ≈194 cm$^{-1}$, corresponding to $B_{3g}$ and $A_g(2)$ modes, respectively. It is observed that the peak positions feature blueshifts compared to their bulk counterpart. The estimated shifts are ≈5 and ≈2 cm$^{-1}$ for $B_{3g}$ and $A_g(2)$ modes, respectively. The observation of blueshift in Raman modes is due to the reduction of the dielectric screening effect upon thinning the bulk crystal layer. These spectral shifts in phonon modes are consistent with our previous results of monolayer SnS sheet,[15] affirming the presence of isolated monolayer SnS sheets.

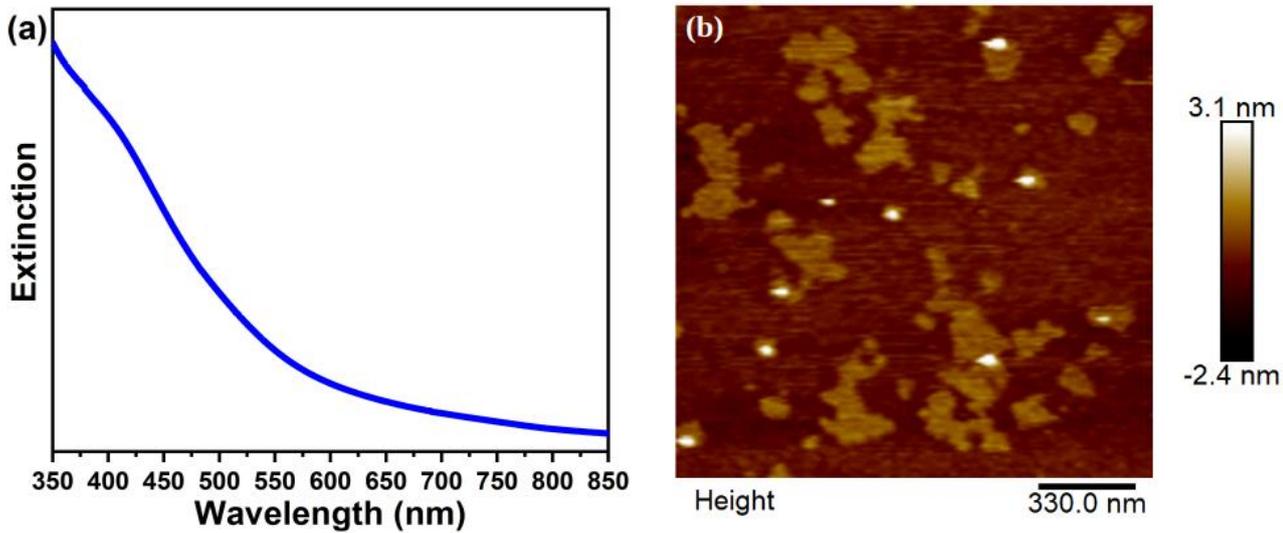

**Figure S12.** a) UV-Vis extinction spectra of isolated SnS sheets (in acetone). b) AFM topography image of SnS sheets. The scale bar illustrates 330 nm.

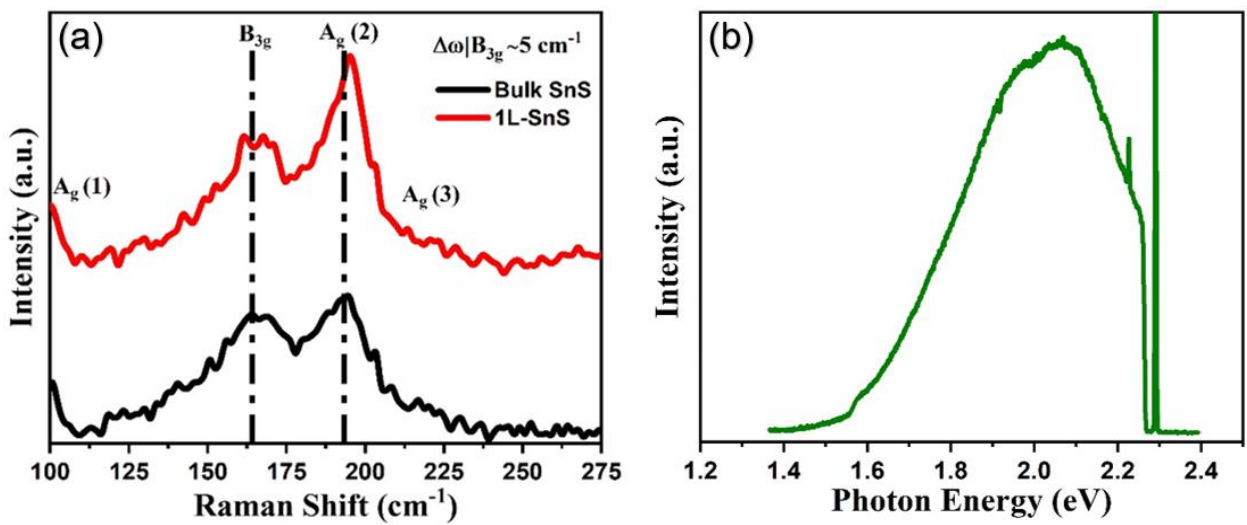

**Figure S13**. a) Raman spectra of bulk and monolayer SnS. b) μ-PL spectra of isolated SnS monolayers. The spectra are collected from the cluster/ensemble of the SnS monolayers in reflection mode. The laser excitation source is 543 nm (2.28 eV), with laser spot size ≈1 μm and incident laser power 640 μW.

**S9.4. Experimental setups for material characterization**

**Extinction spectroscopy**

Extinction spectra of the isolated SnS dispersions are acquired with a Perkin Elmer, Lambda 950 UV/Vis/NIR spectrometer.

**AFM**

Surface topography images of isolated SnS sheets are captured with tapping mode AFM (Multimode AFM from Digital Instruments, Bruker), with controller Nanoscope IIIa. The sample is prepared by drop casting isolated solution on a cleaned $SiO_2$ substrate. All the samples are dried and placed on the microscope stage to scan.

**Raman spectroscopy**

A Nicolet Almega XR μ-Raman analysis system (Thermo Scientific Instruments) is used to collect the Raman spectra. All the spectra are collected with a backscattering geometry, with excitation laser source of 473 nm. The laser beam is focused with a 50× long working distance objective lens with 1800 g(mm)$^{-1}$ grating. The equipment is calibrated with $SiO_2$ before collecting the sample spectra.

**PL spectroscopy**

A home-built backscattering μ-PL setup is used to collect the PL spectra. A laser source of 543 nm (2.28 eV) is used to excite the sample, and a 50× Mitutoyo objective lens focuses the beam to ≈1 μm spot size. The sample is prepared with the drop casting method on $SiO_2$ substrate and dried under vacuum. All measurements are performed under atmospheric pressure.

**S10. Videos**

**Description of video 1**

We present a video of the experimental P-THG imaging of the ultrathin SnS crystals which belong to the field of view shown in Figure 3. We record images of $I_\parallel^{3\omega}$ while rotating the orientation $\varphi$ of the linear polarization of the laser, with $\varphi \in [0°, 360°)$ with step of 4°.

**Description of video 2**

We present a video of the experimental P-THG imaging of the ultrathin SnS crystals which belong to the field of view shown in Figure 3. We record images of $I_\perp^{3\omega}$ while rotating the orientation $\varphi$ of the linear polarization of the laser, with $\varphi \in [0°, 360°)$ with step of 4°.

**Data Availability Statement**

The data that support the findings of this study are available from the corresponding author upon reasonable request.